\documentclass{ifacconf}
\usepackage{graphicx}
\graphicspath{{figures/}}
\usepackage{natbib}
\usepackage{textcomp}
\usepackage{multirow}
\usepackage{float}
\floatstyle{ruled}
\usepackage{gensymb}
\usepackage{tikz}
\usepackage{makebox}
\usepackage{stfloats}
\usepackage{mathtools,cuted}
\usepackage[version=3]{mhchem}
\usepackage{breqn}
\usepackage{algorithm}% http://ctan.org/pkg/algorithms
\usepackage{algpseudocode}% http://ctan.org/pkg/algorithmicx
\usepackage{chngcntr}
\usepackage{enumitem}
\usepackage{xcolor}
\counterwithout{figure}{section}

\begin{document}
	\begin{frontmatter}
		\title{Competing Advanced Process Control via an Industrial Automation Cloud Platform}
		
%		\author[UP, LON]{J. J. Burchell\thanksref{footnoteinfo1}}
		\author[UP]{L.L. Rokebrand}
		\author{J. J. Burchell$ ^{*, **}$}
		\author{L. E. Olivier$ ^{*, ***}$}	
		\author[UP]{I. K. Craig}

		\address[UP]{Department of Electrical, Electronic and Computer Engineering, University of Pretoria, Pretoria, South Africa.}
		\address[SIB]{Sibanye-Stillwater Platinum, Marikana, South Africa.}
		\address[Moyo]{Moyo Africa, Centurion, South Africa.}
		
		\begin{abstract} % Abstract of not more than 250 words.
				This paper proposes an innovative approach for the advanced control of an industrial process via an automation cloud platform. Increased digital transformation and advances in Industrial Internet of Things (IIoT) technologies make it possible for multiple vendors to compete to control an industrial process. An industrial automation cloud platform facilitates the interaction between advanced process control (APC) vendors and the process. A selector, which forms part of the platform, is used to determine the best controller for a process for any given time period. 
The article starts with a general overview of platform businesses, platforms aimed at industry, and the steps required to build such platforms. Issues that need to be addressed to make APC via an automation platform practically viable are discussed including what process information to provide to APC vendors, continuous evaluation of controllers even when not in control of the process, bumpless transfer, closed-loop stability, constraint handling, and platform security and trust.
A case study is given of competing APCs via an industrial automation cloud platform. The process used in the study is a surge tank from a bulk tailings treatment plant, the aim of which is to keep the density of the tank outflow constant while maintaining a steady tank level. A platform facilitates the competition of three vendors for control of this process. It is shown that the cloud platform approach can provide the plant access to a superior controller without the need for directly procuring the services of an exclusive vendor.
	 
		\end{abstract}
		

\end{frontmatter}

%===============================================================================
%///////////////////////////////////////////////////////////////////////////////
\section{Introduction}\label{Section:Introduction}

Platform businesses have come to dominate the Internet and are playing an ever larger role in the global economy. Seven of the top ten companies in the world today (measured in terms of market capitalisation) are platform businesses \citep{pwc}. Traditional businesses follow a linear business model in which they add value to “raw materials” by creating products/services that are sold on to consumers. In contrast, a platform business facilitates the exchange of value between two or more user groups \citep{Moazed2016}. Consumers and producers are brought together so they can interact in unique ways, with the potential for exponential increases in utility and value \citep{Cusumano2019}. Traditional linear businesses create value within the boundaries of a company or a supply chain, whereas platform businesses create value by utilizing an ecosystem of producers, complementary products and consumers \citep{Hein2020}.

Platform businesses can be classified into two broad categories, i.e. transaction or exchange platforms, and innovation or maker platforms (\cite{Cusumano2019}; \cite{Moazed2016}). Examples of transaction platforms include Alibaba, LinkedIn and Uber, and innovation platforms include Apple iOS, Google Android and Intel CPU. Some companies have both transaction and innovation components and are referred to as hybrid platform businesses \citep{Cusumano2019}. Well known hybrid platform businesses include Apple, Microsoft and Tencent. As the name suggests, transaction platforms provide value primarily by optimizing exchanges directly between consumers and producers, whereas innovation platforms generate value by enabling producers to create complementary products and distribute them to a large number of consumers (\cite{Cusumano2019}; \cite{Moazed2016}).

The above mentioned businesses typically have a consumer focus and as such are well known to the general public. There are however an increasing number of platforms that focus on industry, and the race is on to see which company will best utilize Industrial Internet of Things (IIoT) technologies to build the dominant IIoT cloud platform \citep{Farnell}. The global market for IIoT-enabled business models in the industrial equipment and machinery space is expected to grow substantially, and linking industrial automation and IIoT platforms is considered the industry's new frontier \citep{Bolz2018}.

Industrial automation vendors worldwide are undergoing a shift to software and managed services, and many of the top 50 global automation vendors now have their own IIoT cloud platforms (\cite{Isaksson2018}; \cite{O'Brien2019}). Well-known examples include ABB Ability, GE Predix, Honeywell Sentience, and Siemens MindSphere. These platforms have the potential to provide increasing value by enabling exchange between industrial equipment, machinery, producers, complementary products and consumers. Current industrial automation platforms are designed primarily for one-way communication, where the main aim is to get user data into the cloud. Once in the cloud, value is added to the data by providing e.g. visualization tools, asset performance management solutions, and remote monitoring (\cite{ABB2017}; \cite{RealiSCADA}). 

Closing the loop through the cloud is currently limited in the industrial automation industry mainly to loops that are not operationally critical such as higher level production optimization loops. Internet-based process control is however not a new topic (see e.g. \cite{Yang2003}), and numerous authors have investigated the impact of closing lower level control loops via the cloud (see e.g. \cite{Givehchi2014}; \cite{Hegazy2017}; \cite{Xia2012}; \cite{Xia2015}).  Although there are some reported examples of Model Predictive Control (MPC) over an IIoT platform (see e.g. \cite{Nunez2020}), closing the loop at the supervisory control level via an industrial automation cloud platform as proposed here is not current practice. It is certainly possible, and the aim of this paper is to show that it can be beneficial as well.

Industrial companies have been outsourcing their  IT (Information Technology) functions for many years, but OT (Operation Technology) functions, being part of the core business, are usually performed in-house. It is therefore understandable that companies might be reluctant to outsource some of their OT functions to the cloud as proposed in this paper.  Corporate IT services are however increasingly being performed in the cloud, and with the growing convergence between OT and IT \citep{Klaess2019}, one would expect that organisational cultures will evolve over time such that cloud-based OT functions, such as higher level automation functions, will become more common.

This paper proposes an innovative approach for the advanced control of an industrial process via an automation cloud platform. It shows how multiple vendors can compete, potentially in real-time, to control an industrial process at the supervisory control level. An industrial automation cloud platform facilitates the interaction between multiple APC vendors and the process. A selector, which forms part of the platform, is used to determine the best controller for a process for any given time period. This approach can allow for increased competition in the industrial automation industry, and is illustrated here using a case study in which a surge tank from a bulk tailings treatment plant is controlled. The work presented in this paper expands on \cite{Craig2020}. 

An overview of competing APCs via a cloud platform is given in Section \ref{Section:Platform Businesses} together with a brief description of the steps required to build a successful platform business. A discussion about the practical considerations surrounding the proposed platform is presented in Section \ref{Section:Platform}. The dynamic model of the system is derived in Section \ref{Section:DynamicModelling} and an input-output controllability analysis is performed on the model in Appendix \ref{Appendix: IO cntrl}. Section \ref{Section:Controllers} presents three different APCs employed by different vendors, one conventional feedback and two MPC controllers as well as a rudimentary local fallback controller. Section \ref{Section:Competing} presents the scenario where the three APCs compete to control the process with the selector selecting the one it deems the best.

% While some works explored the possibility of moving the actual controller to a cloud based service none have proposed an access economy platform which would facilitate the crow-sourcing of control solutions to plants in real-time (\cite{Givehchi2014}; \cite{Hegazy2017}; \cite{Xia2012}; \cite{Xia2015}). The authors are of the opinion that the success of access economy and crowd-sourcing business models being employed by large corporations in many different sectors can achieve similar success in the field of industrial automation and process control \citep{topcoder}.  

\section{Platform Businesses} \label{Section:Platform Businesses}

\subsection{Competing APCs via a Cloud Platform} \label{sec: Platform Overview}

\begin{figure}[h!]
	\centering
	\includegraphics[width=9 cm]{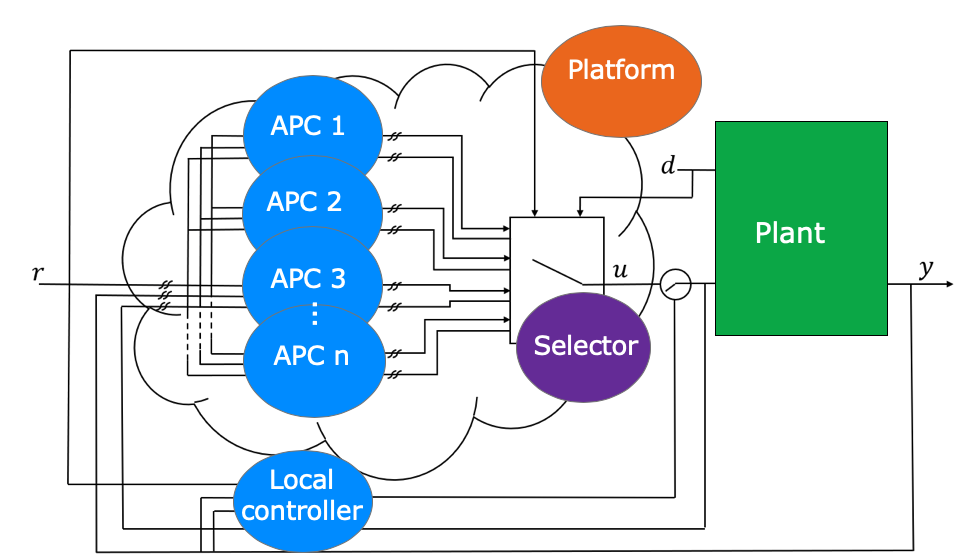}
	\caption{Competing APC controllers via a cloud platform.}
	\label{fig:Competing_APC}
\end{figure}

A graphical description of competing APCs via a cloud platform is given in Fig.~\ref{fig:Competing_APC}. Multiple vendors compete to control an industrial process and their APC solutions reside in the platform as applications (apps). A selector, which also forms part of the platform, evaluates the different APC controllers over a period of time, and at the end of each period, selects and implements the controller which has been determined optimal based on some type of performance measure. It is envisaged that the selector would be owned by the platform and be configurable by the users of the APC technology. A local fallback controller resides on-site and is used in case of a problem such as a broken internet connection.

\subsection{Which Type of Platform Business?} \label{sec: Platform Type}

Which type of platform business, transaction, innovation or hybrid, is best suited to the process automation industry? It is proposed that the best initial fit for an industrial automation platform is an innovation-type platform. Such a platform would aim to entice and enable third-parties to create complementary products or apps on the platform in order to populate the platform ecosystem \citep{Hein2020}. The result will be an app store for industrial automation users similar to e.g. the Apple app store, the Google Play store and the Tencent My app store. Once successful, an industrial automation platform can later evolve into a hybrid platform by introducing e.g. transaction components such as the Digital Marketplace which is part of Mindsphere \citep{Mindsphere}.

\subsection{Steps Required to Build a Successful Platform Business}\label{sec: Platform Steps}

There are a number of steps that need to be followed in order for an automation platform to be successful \citep{Cusumano2019}. These include:
\begin{enumerate}
	\item Building your own platform, buying a platform or belonging to someone else's platform.
	\item Choosing platform market sides, e.g. users and suppliers of automation technology.
	\item Solving the chicken-or-egg problem, i.e. which market side should initially be focused on?
	\item Designing a business model, i.e. how will the platform make money?
	\item Ecosystem governance. 
\end{enumerate}

\subsubsection{Build, buy or belong}\

The are a number of approaches to developing an IIoT platform \citep{Phillips2018}. \cite{Cusumano2019} have identified three main approaches: building your own platform; buying a platform; or belonging to someone else's platform. Building your own platform can be a very expensive undertaking. GE Predix has not lived up to expectation despite GE having spent billions of dollars on its development  \citep{NYTimes_2018}. Major industrial automation vendors have tied up with cloud service providers to develop their application on.  For example, Siemens' MindSphere was originally based on the SAP cloud platform, and has now also been released on Amazon Web Services, Microsoft Azure and Alibaba Cloud \citep{Mindsphere}.

\subsubsection{Choosing platform market sides}\

\begin{figure}[h!]
	\centering
	\includegraphics[width=9 cm]{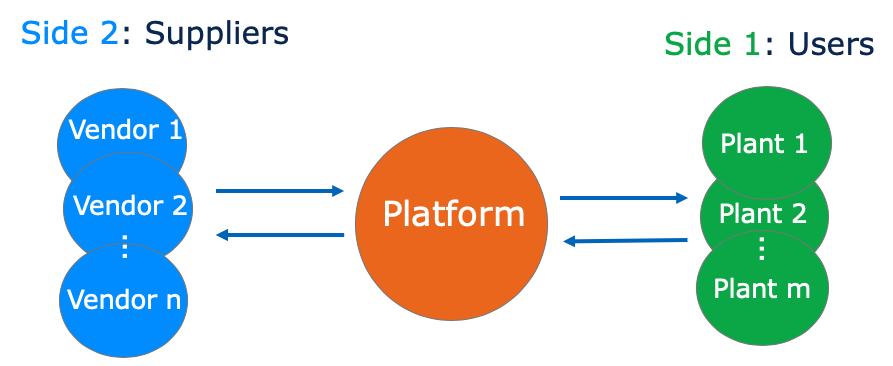}
	\caption{Platform market sides.}
	\label{fig:Platform_market_sides}
\end{figure}

The idea is to start simple. Two market sides are sufficient for a start. Market side 1 will typically be the users of automation hardware and software, and side 2 the suppliers thereof as shown in Fig.~\ref{fig:Platform_market_sides}.

It is important early on to determine the value proposition for each market side as well as for the platform itself. Value for users include easy access to products and skills not available in-house, and lower cost of technology owing to increased competition. Suppliers can gain access to a wide range of users at low cost, and can roll out solutions to many users simultaneously and receive almost instant feedback. By reducing transaction cost and enabling complementary innovations, the platform can scale relatively quickly which potentially allows for exponential growth in value.

\subsubsection{Solving the chicken-or-egg problem}\

How should the platform business start? The attractiveness of the platform for users depend on the number of supplier apps, and the number of suppliers willing to develop apps for the platform depends on the number of users on the platform \citep{Ruutu2017}. A possible way to start is for the platform owner to first create stand-alone value for users by developing its own apps. Such apps can include data visualization tools, condition monitoring and predictive analytics. Once enough users are on board, third-party suppliers could be enticed to create apps for the platform. This is then also the route that the major automation vendors have taken with the development of their platforms. The initial focus is therefore on getting as much user equipment and operational data into the platform as possible, even if users are initially subsidised to get their data into the owner's cloud platform. 

An increasing number of users will attract additional suppliers. The eventual aim is to get as many third-party apps on the platform as possible. It should therefore be made easy for suppliers to develop apps for the platform by the owner providing e.g. inexpensive application programming interfaces and software development kits. 

\subsubsection{Designing a business model}\

How will the platform make money and eventually become profitable? Scale usually has to be created before a platform business can become profitable. Revenue growth on its own however does not guarantee profitability. A well known example of a transaction platform business that struggles to make a profit, despite its scale, is the ride-hailing company Uber \citep{Reuters2020}.

Platform businesses often initially subsidise users to build up one side of the market. This can be a costly but necessary exercise in order to scale as quickly as possible. Once the platform is creating sufficient value, ways have to be found in which the value created can be monetised. Examples include getting users to start paying for the platform and apps, and/or getting a portion of the sale of complementary products sold on the platform.

Winner takes all (or most) platforms have emerged in the mobile phone industry and internet search \citep{Ruutu2017}. It is too early to tell if such winning platforms will emerge in the industrial automation industry. Automation vendors are facing competition in the IIoT field not just from each other, but also from new entrants that include cloud suppliers such as Amazon and Microsoft, business software companies such as IBM and SAP, and start-ups such as C3 IoT and Uptake \citep{NYTimes_2018}.

 \subsubsection{Ecosystem governance}\
 
 A platform together with its complements and market sides are called an ecosystem as there is mutual dependence between all participants. Good governance of platform ecosystems is crucial in order to build trust and maintain quality \citep{Schreieck2017}. In this sense good governance refers to e.g. what is allowed and encouraged, what is forbidden and discouraged, and how rules will be enforced.  The platform owner also has to decide on the degree of openness - a balance has to e.g. be found between scaling as quick as possible and maintaining quality \citep{Ruutu2017}. Platform companies have been known to compete directly with their most successful complementors, and ecosystem participants would want to know where they stand in this regard \citep{Wen2019}.

In summary, a successful automation platform should provide an operating system for the IIoT with open standards and interfaces to facilitate the upload of user equipment and operational data. In addition, it should provide a home for data from its own and third-party equipment vendors, as well as an open ecosystem that houses its own and third-party developer apps.

%///////////////////////////////////////////////////////////////////////////////

\section{Facilitating Platform for Industrial Control} \label{Section:Platform}

There are a number of issues that will need to be addressed and clarified in order for competing APCs via a cloud platform to become practically viable. These include:
\begin{enumerate}
	\item In order to evaluate the performance of the competing controllers, it would be necessary to simulate the control actions provided by each controller on a model of the plant over a period of time. 
	\item A process model would thus be needed by the selector for the evaluation of each controller, as well as by each competing APC vendor for controller development. 	
	\item Some kind of bumpless transfer mechanism may need to be employed when switching takes place.
	\item Constraints on the manipulated variables (MVs) and controlled variables (CVs) will need to be communicated to the competing controllers, as well as the selector.
	\item It would be necessary to have a local controller to serve as a fallback strategy in the event of a technical fault.	
	\item Overall stability of the system would need to be addressed.
	\item Cyber security and trust in the platform need to be addressed.   
\end{enumerate}

These issues are discussed in more detail in the following subsections.

\subsection{System Identification and Modelling} \label{sec: modelling}

System identification and modelling typically makes up a large portion of the development of an APC solution such as an MPC \citep{Isaksson2018}. This would typically include modelling the system dynamics as well as determining all the relevant plant specific information such as actuator and sensor capabilities, constraints and models \citep{Forbes2015}. 

Typical modelling procedures include deriving transfer functions from step test data \citep{Seborg2011} or using subspace identification to yield a state-space model \citep{Qin2006}. It is possible that each competing vendor could develop their own model from input-output data. The selector will use the most accurate model available to evaluate the performance of the competing APC controllers. This would thus involve supplying each controller with the outputs $\mathbf{y}$ of the plant as well as the control actions $\mathbf{u}$ implemented on the plant (see Fig.~\ref{fig:Competing_APC}). Although this may be the most viable approach in practice, it would however be ideal that a validated first-principles state-space model such as described in \cite{Wiid2020} be made available to the selector and all competing APC controllers.

Relevant sensor and actuator information would need to be made available to the controllers, which would include all constraints on the MVs and CVs such as the upper and lower saturation levels as well as any rate of change limits. It would also be necessary for all constraints on the MVs to be made available to the selector in order to ensure that no attempt is made to implement any unachievable control actions on the plant.

As mentioned previously, if information such as the relevant MVs, CVs, constraints and plant model are known, this is already a large part of developing an APC solution completed. Obtaining this information would require significant process knowledge and plant access. Once this step is complete there are many options and solutions one could employ to develop an APC, thus processes which are common and well defined would be best suited to a competing APC cloud platform approach \citep{Darby2012}. A vendor could be employed to assist with the problem definition and modelling of a process. Alternatively, an on-demand remote workforce approach can be used drawing from the vast pool of experience and expertise which could be available from a crowd-sourcing application -- see e.g. \citep{topcoder}.

It is often the case in practice that much of the process information, particularly the measured output data, would be considered as sensitive and confidential. In this case all values made available by the plant (including measurement data, constraints and the plant model) could simply be scaled. It is also possible that potential vendors be vetted on non-critical or non intellectual property (IP) sensitive processes (or even on a simulation if that is preferable). Although this would still render the proposed platform feasible, a successful IIoT platform involves a much greater level of openness and access to information than more traditional business models \citep{Ruutu2017}. The proposed platform would thus have a greater potential for improving system performance if as much information as possible about the plant was shared, allowing for as much expertise as possible to be attracted to the proposition of supplying an APC solution.  

\subsection{Controller Performance Measurement} \label{sec: performance measure}
  
In order to evaluate the performance of the competing APCs, it would be necessary for the selector to simulate each controller in closed-loop on a model of the plant. In terms of following the reference $\mathbf{r}$ (see Fig.~\ref{fig:Competing_APC}), this process is straight forward so long as the desired reference is supplied to the selector. Determining the disturbance rejection capabilities of the competing controllers will however require knowledge of the actual process disturbances -- $\mathbf{d}$ as given in Fig.~\ref{fig:Competing_APC}. If the disturbances can be measured, they could readily be supplied to the selector, whereas if they are not, an estimate thereof would need to be determined by the selector through the use of an observer.

Some type of performance measure will be needed to determine the best controller based on the simulated outputs and generated control moves of each controller. One such performance index could be the sum of the squares of the error ($SSE$) in the simulated outputs for each evaluation horizon:
\begin{equation}
\label{eq:min sse}
J_i=\sum_{j=k-N_e+1}^{k}(\mathbf{r}(j)-\mathbf{y}_i(j))^T\mathbf{W}_e(\mathbf{r}(j)-\mathbf{y}_i(j))
\end{equation}
where $\mathbf{y}_i$ is the simulated output of the $i^{th}$ controller, $\mathbf{W}_e$ is a weighting matrix which is used to place more or less emphasis on the different outputs and $N_e$ is an evaluation horizon or time period over which the controller is simulated. 

The performance measure in (\ref{eq:min sse}) would typically favour a more aggressive controller due to the exclusive penalisation of the deviation of the simulated outputs $\mathbf{y}_i$ from the reference $\mathbf{r}$. A slightly modified version could be represented as:
\begin{equation}
\label{eq:min sse_du}
\begin{split}
J_{i}=\sum_{j=k-N_e+1}^{k}&(\mathbf{r}(j)-\mathbf{y}_i(j))^T\mathbf{W}_e(\mathbf{r}(j)-\mathbf{y}_i(j))\\
		& + \Delta \mathbf{u}_i^T(j)\mathbf{W}_u 	\Delta \mathbf{u}_i(j)
\end{split}
\end{equation} 
where $\mathbf{u}_i$ is the control move generated by the $i^{th}$ controller,  $\Delta \mathbf{u}_i(j) = \mathbf{u}_i(j) - \mathbf{u}_i(j-1)$ and $\mathbf{W}_u$ is a weighting matrix used to place more or less emphasis on the utilisation of the respective control actions. This measure would also penalise excessive control action and therefore not favour a more aggressive controller. 

The performance measure and relevant weighting factors such as $\mathbf{W}_e$ and $\mathbf{W}_u$ would ideally be known to the vendors that supply the APCs, with this decision possibly lying in the hands of the plant or the selector. With vendors given the same plant model and objective function as in (\ref{eq:min sse_du}), one would expect their APCs solutions to display similar performance. Factors that might distinguish one solution from the other could however include superior disturbance rejection properties and/or controller cost.  Again, from a cloud platform business point of view, the more knowledge the vendors have regarding the desired performance specifications of the plant, the better they would be able to tailor their proposed solutions to these requirements. 

The proposed manner of switching is for the selector to simulate the controllers on the plant model, in parallel with the running plant using the actual reference signal $\mathbf{r}$ and disturbance $\mathbf{d}$ (measured or estimated) over a particular evaluation horizon $N_e$. At the end of each evaluation horizon, a performance measure such as (\ref{eq:min sse}) or (\ref{eq:min sse_du}) would be calculated based on the simulated outputs and generated control moves of each APC controller. The controller which achieves the best value for the performance index will then automatically be selected and control of the actual plant switched over to this controller. The performance indices for each controller would then be set to zero and the process repeated. 

It is worth noting that other manners of switching are possible with the proposed platform philosophy. For example, the decision to switch can lie with the plant, with the controllers being simulated by the selector in advance on predetermined reference and disturbance signals $\mathbf{r}$ and $\mathbf{d}$ which are representative of the expected future conditions of the system. This would allow for the best APC controller to be selected in advance in a predictive manner.

The main goal of the platform is to facilitate the procurement of the best available APC option given the operating conditions of the plant and the specified performance requirements.  `Best' could also refer to the price of the APC in addition to superior setpoint following and limited control moves, if the selector performance index (\ref{eq:min sse_du}) is amended to include a price component. 

\subsection{Bumpless Transfer} \label{sec:bumpless transfer}

In the event that one of the competing APCs is determined to be optimal and is to be selected, it will likely be the case that the control moves generated by this controller in simulation with the plant model are not equal to the current control moves implemented on the actual plant, but may in fact differ substantially. This will require a bumpless transfer mechanism to be employed by each controller in the event that it is awarded control of the actual plant. 

In order to achieve this, knowledge of the present control moves implemented on the plant at the time when switching is to take place will be required by the new controller. This responsibility would inherently fall with the individual controllers and would be achieved in different manners depending on the type of controller. For illustrative purposes, two methods are proposed: one for MPC controllers and the other for conventional feedback controllers. Both these methods are used in the case study described in Section \ref{sec: CaseStudy}.

For MPC controllers using the state-space formulation of \cite{Mayne2009}, bumpless transfer can be achieved by employing two observers for state/disturbance estimation. One would estimate the states and disturbances based on the measurement of the actual plant output and the actual control action implemented on the plant (both pieces of information which are available to the controllers), and a second would provide estimates based on the model it is controlling. When the MPC controller is not selected, the estimate derived from the simulated plant model and control signal are used in the MPC optimisation algorithm. If the MPC controller is selected, the estimate derived from the actual plant output and control signal (which will now become the control signal generated by this controller) is used in the MPC optimisation algorithm and seamless transfer of control takes place. 

The conventional feedback controllers would employ a different principle in which the integral term of the controller is back initialised to yield a control action equal to that of the previous control action applied to the plant. Any conventional, linear feedback controller can be represented in state-space form as:
\begin{equation}
\label{eq:cntrl_ss}
\dot{\mathbf{x}}_{ci} = \mathbf{A}_{ci}\mathbf{x}_{ci}+ \mathbf{B}_{ci}\mathbf{e}_i 
\end{equation}
\begin{equation}
\mathbf{u}_i = \mathbf{C}_{ci}\mathbf{x}_{ci} + \mathbf{D}_{ci}\mathbf{e}_i 
\end{equation}
where $\mathbf{A}_{ci}$, $\mathbf{B}_{ci}$, $\mathbf{C}_{ci}$ and $\mathbf{D}_{ci}$ are the state-space coefficient matrices for the $i^{th}$ controller, $\mathbf{e}_i$, $\mathbf{u}_i$ are the simulated error and control signals respectively and $\mathbf{x}_{ci}$ is the internal state which incorporates the integral action. Each state-space system can then be transformed into a discrete state-space system as:
\begin{equation}
\label{eq:cntrl_ss_d1}
\mathbf{x}_{ci}(k+1) = \boldsymbol{\Phi}_{ci}\mathbf{x}_{ci}(k)+ \boldsymbol{\Gamma}_{ci}\mathbf{e}_i(k) 
\end{equation}
\begin{equation}
\label{eq:cntrl_ss_d2}
\mathbf{u}_i(k) = \mathbf{C}_{ci}\mathbf{x}_{ci}(k) + \mathbf{D}_{ci}\mathbf{e}_i(k) 
\end{equation}  
where the discrete coefficient matrices $\boldsymbol{\Phi}$ and $\boldsymbol{\Gamma}$ are determined as:
\begin{equation}
\label{eq:Phi-discrt}
\boldsymbol{\Phi} = e^{\mathbf{A}\Delta t}
\end{equation}
\begin{equation}
\label{eq:Gam-discrt}
\boldsymbol{\Gamma} = \int_0^{\Delta t}e^{\mathbf{A}s}ds \cdot \mathbf{B} = \mathbf{A}^{-1}\cdot [e^{\mathbf{A}\Delta t} - \mathbf{I}] \cdot \mathbf{B}
\end{equation}
where $\Delta t$ is the sampling period. In order to back initialise the controllers, it follows from (\ref{eq:cntrl_ss_d2}):
\begin{equation}
\label{eq:cntrl_ss_init}
\mathbf{x}_{ci}(k) = \mathbf{C}_{ci}^{\dagger}(\mathbf{u}(k-1) + \mathbf{D}_{ci}\mathbf{e}(k))
\end{equation} 
where $\mathbf{C}_{ci}^{\dagger}$ is the pseudoinverse of the matrix $\mathbf{C}_{ci}$, $\mathbf{u}(k-1) $ is the previous control action applied to the plant and $\mathbf{e}(k)$ is the actual error based on the measured output of the plant. This method will successfully initialise the controllers to output a control action which is representative of the current state of the plant and prevents the controllers from generating any excessive control action in the event that control is switched over to them.  

\subsection{Constraint Handling} \label{sec: constraint handling}

Constraints on the MVs and CVs will need to be communicated to the competing APCs and the selector. This information is required for controller development, and to prevent the implementation of control moves which violate the constraints. Constraint handling is typically incorporated into most MPC formulations as part of the optimisation problem, but there is no explicit constraint handling present in the conventional feedback controller form. This would have to be implemented ad-hoc by each conventional feedback controller by setting an output equal to its given constraint in the event that the output generated by the controller exceeds this constraint. If this is the case, the $\boldsymbol{\Gamma}_{ci}$ matrix of the controller in (\ref{eq:cntrl_ss_d1}) is set to zero to disable any integral action and thus prevent any integral windup. Other more sophisticated anti-reset windup schemes may also be employed \citep{Seborg2011}.

The selector would also assist in this matter by restricting the plant inputs to their relevant constraints should the active APC attempt to violate them. This would also be true for any rate of change constraints on the MVs. Should this happen, it would be possible to flag a controller and incorporate this information in future switching decisions, and if a controller violated any of the MV constraints during the simulation over an evaluation horizon, it could be automatically excluded from selection at the end of that particular horizon.

\subsection{Fallback strategy} \label{sec: fallback strategy}

With the competing APCs possibly all being in a remote location, it would be necessary to have a local controller (see Fig.~\ref{fig:Competing_APC}) to serve as a fallback strategy in the event of a technical fault with a sensor or actuator, a communication fault with the selector or between the selector and active controller, or some undesirable control actions from the active controller. By the very nature of the requirements placed on this controller, it is necessary that it be locally installed at the plant. It is also assumed that all base layer control would be pre-existing and locally installed.

Many plants do not have on-site personnel that are able to design, install and maintain APC solutions, and as a result such solutions are usually procured from APC vendors. The local fallback controller would thus typically be a relatively rudimentary controller (such as a heuristically tuned PI controller) implemented by plant personnel. The goal of the platform is to thus facilitate a new method of procuring an APC which is superior in performance to the local fallback controller.  

\subsection{System Stability} \label{subsec: stability}

The overall closed-loop system stability would need to be addressed, as switching in a dynamic system can be a potential source of instability \citep{Liberzon2003}. It is known from the switched systems literature that switching between controllers which yield individually stable closed loop systems can in fact cause the overall switched system to be unstable (\cite{Liberzon2003}; \cite{Astrom1996}). It has also been shown however, that if the individual controllers all yield stable closed loop systems, the switched system will be stable as long as intervals between switching are longer than the corresponding slowest individual closed loop system dynamics \citep{Leith2003}. 

For the purposes of the proposed platform this translates into a minimum bandwidth requirement for each competing controller. This would need to be determined and specified by the plant from the process dynamics and input constraints and communicated to the controllers via the selector. It would seem counter intuitive that a very sluggish controller would be determined optimal and thus selected, but the actual transient response dynamics may need to be verified by the selector before selecting the controller. 

It is worth noting that switching at intervals longer that the slowest expected closed loop dynamics is a sufficient condition for stability (given all individual systems are stable) but not a necessary condition \citep{Lin2009}. To guarantee stability otherwise would require knowledge of the individual controller realisations, which would not be the case for the platform scenario, thus switching at longer intervals is necessary to guarantee stability. 

\subsection{Platform Security and Trust} \label{subsec: security}

In the proposed platform philosophy, where potentially any willing party may be awarded control (and hence access to local control loops) of the plant, there would need to be a certain amount of trust in the APC vendors, or at least some level of accountability. The level of security and trust required by a plant may differ from process to process depending on safety concerns, the impact of mistakes, and IP, among others.

The security and IoT reputation of cloud platforms wanting to perform OT functions could first be assessed using e.g. a trust assessment framework proposed by \cite{Li2019}. Additional trust building mechanisms that could be employed include reputation, measurement and rating systems; and cloud transparency mechanisms and formal accreditation \citep{Werff2019}. The APCs need to prove their ability to control the plant by the selector evaluating their performance on a model of the plant, but this may not prevent a malicious agent from being awarded control of the plant and subsequently attempting to sabotage the process if selected. If this is the case, or if an APC exhibits any undesirable behaviour, the relevant vendor could be penalised via e.g. a rating system. Controllers which are frequently selected and never exhibit any undesirable behaviour would subsequently improve their rating. Plants which require a high level of trust could then specify a certain minimum performance rating for APCs allowed to compete to control their process.  

Similar to the issue of sabotage is that of cyber security, as it may be possible for malicious agents to hack into the platform in order to exploit the weaknesses created in the gaps that still exist between IT and OT \citep{O'Brien2019}. Cyber security is taken very seriously by the major automation vendors as this is key to the success of their IIoT cloud platforms. For example, ABB Ability offers a cyber network security management service to users \citep{ABBCyberSecurity}. IIoT cloud platforms owners could initially subsidise users to strengthen their cyber security in order to build up the user side of the market as discussed in Section \ref{Section:Platform Businesses}.

\subsection{Controller Remuneration} \label{subsec: renumeration}

Vendors would need to invest resources into developing an APC strategy. This might not be cost effective for a single solution, but a cloud platform could potentially provide the vendor with access to many similar users to which the same solution could be employed with relatively minor adjustments.  

When APC vendors compete via an industrial automation cloud platform, they will only be remunerated when they are selected to control the plant. This would ideally be dependant on the marginal economic benefit achieved from using the relevant APC as described in \cite{Bauer2008}.  It is however the experience of vendors that users prefer to pay a fixed price for an APC solution partly due to the way that funds for its procurement are motivated for.  Users are however often willing to enter into service level agreements (SLAs) with APC vendors, which is an additional cost that would not be required when using a competing APC platform. The fixed price that users are willing to pay for an APC, together with the cost of an SLA, could act as a price ceiling that users offer to competing APC vendors over a given time period. It is expected however that a competing APC platform would result in prices that are substantially lower than this price ceiling due to increased competition. 

Remuneration could be done via conventional means or even through the use of blockchain technologies such as a crypto currency or smart contract, both of which are receiving increasing amounts of interest in IoT applications (\cite{Huckle2016}; \cite{Rahman2019}). 
  
  %All the aforementioned considerations are incorporated into the schematic in Fig. \ref{fig:competing_controllers_1} and the result is shown in Fig.~\ref{fig:competing_controllers_5} which shows the measured disturbances $\mathbf{d}$ and reference $\mathbf{r}$ supplied to the selector, the feedback channel used to simulate the competing controllers, the implemented control move $\mathbf{u}$ fed back to the controllers $\mathbf{G}_{c1}, \mathbf{G}_{c2}... \mathbf{G}_{cn}$, as well as the local fallback controller $\mathbf{G}_{c\_local}$. 

%\begin{figure}[h!]
%	\centering
%	\includegraphics[width=9 cm]{competing_controllers_5}
%	\caption{Competing controllers with local fallback controller.}
%	\label{fig:competing_controllers_5}
%\end{figure}

\section{Case Study: Competing APC via a Cloud Platform} \label{sec: CaseStudy}

\begin{figure}[h!]
	\centering
	\includegraphics[width=9 cm]{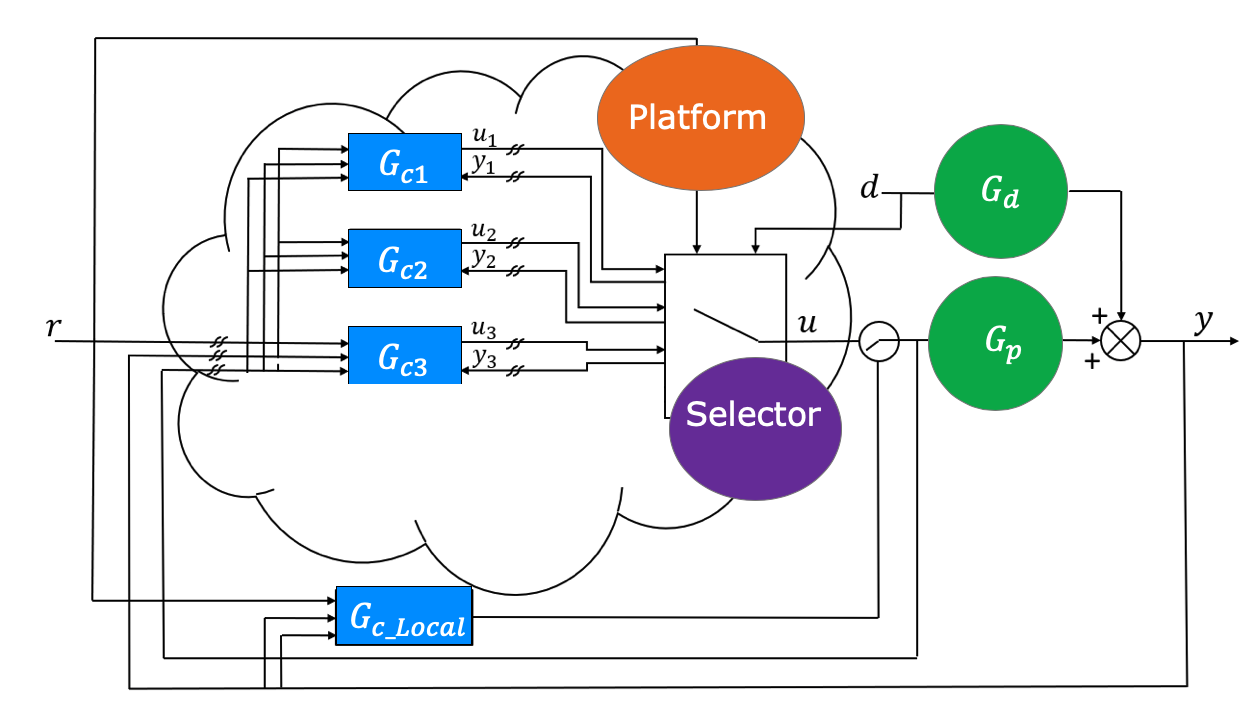}
	\caption{Case Study: Competing APC controllers via a cloud platform.}
	\label{fig:Competing_APC_2}
\end{figure}

This section presents a case study of the proposed platform. There is one user plant that is controlled ($\mathbf{G}_{p}$ and $\mathbf{G}_{d}$ -- an output disturbance is assumed), i.e. the surge tank from a bulk tailings treatment plant. Three vendor APCs ($\mathbf{G}_{c1}$, $\mathbf{G}_{c2}$ and $\mathbf{G}_{c3}$) compete to control the process and a local fallback controller ($\mathbf{G}_{c\_local}$) is available as backup. A graphical depiction of competing APCs via a cloud platform as used in the case study is given in Fig.~\ref{fig:Competing_APC_2}.  

A brief description of the process in question and a derivation of the plant model is presented. This is followed by a description of the local fallback controller and the three vendor APCs competing to control the plant. A description of the platform philosophy as applied to the case study is then presented followed by the results of the platform simulation. 

\subsection{The Plant}\label{Section:DynamicModelling}

The bulk tailings treatment (BTT) plant receives chrome tailings from a tailings dam. The purpose of the BTT plant is to dewater and stabilise the tailings before feeding it to a chrome concentrator which produces metallurgical and chemical grade chrome using spiral concentrators. These spiral concentrators sort the feed into the different chrome grades using gravity separation, with their separation efficiencies sensitive to disturbances in density \citep{HollandBatt1982}. A key objective for the BTT plant is therefore to supply the chrome concentrator with a stable feed density.

\begin{figure*}[h!]
	\centering
	\includegraphics[width=6.2in]{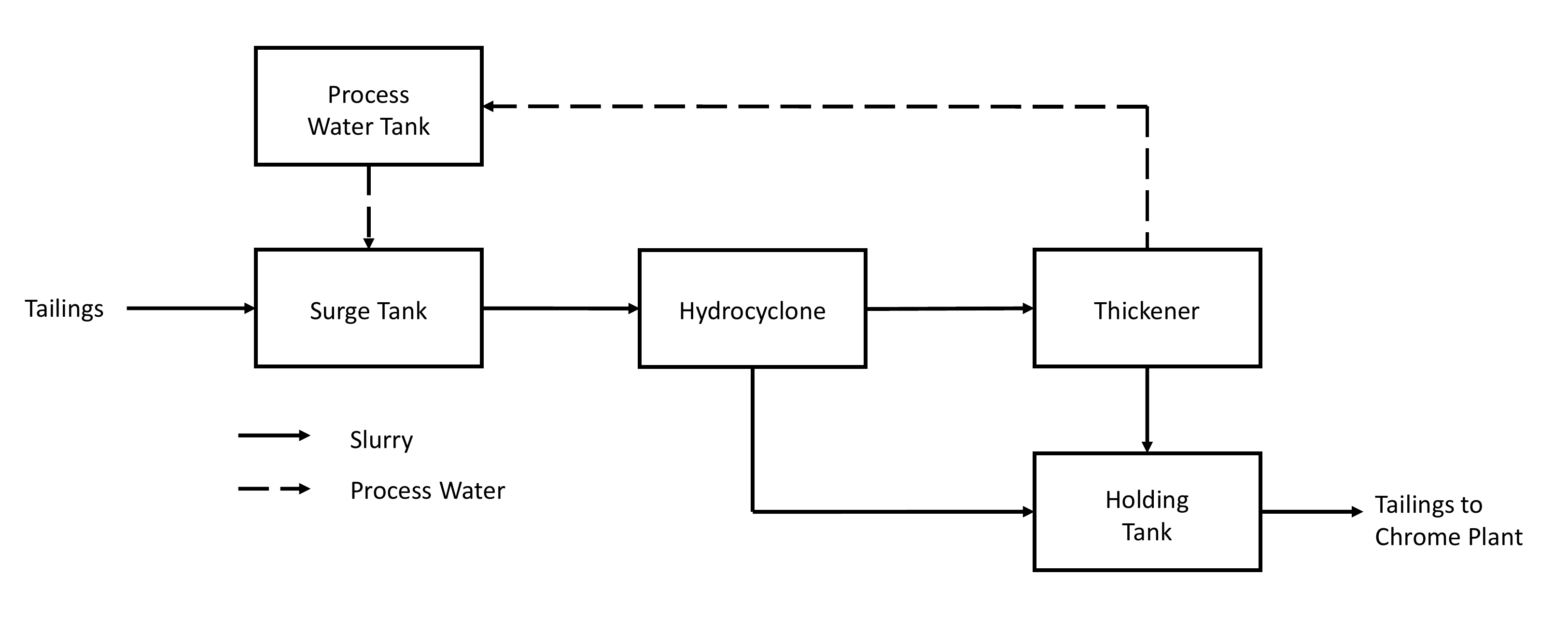}
	\caption{Flowsheet of the Bulk Tailings Treatment (BTT) plant.}
	\label{fig:BTTFlowSheet}
\end{figure*}

An overview of the BTT plant flowsheet is presented in Fig. \ref{fig:BTTFlowSheet}. The hydrocyclone and thickener are used for dewatering, or densification, while the surge tank uses process water to stabilise the incoming tailings feed density. The hydrocyclone feed density has by far the largest impact on hydrocyclone efficiency \citep{Ntengwe2011}. This investigation focuses on stabilising the surge tank density, which would result in compounded improvements in the cyclone and thickener efficiencies. A dynamic model for the surge tank is derived, and used to compare the potential of an APC to improve on the local fallback controller.
\begin{figure}[h!]
	\centering
	\includegraphics[width=8cm]{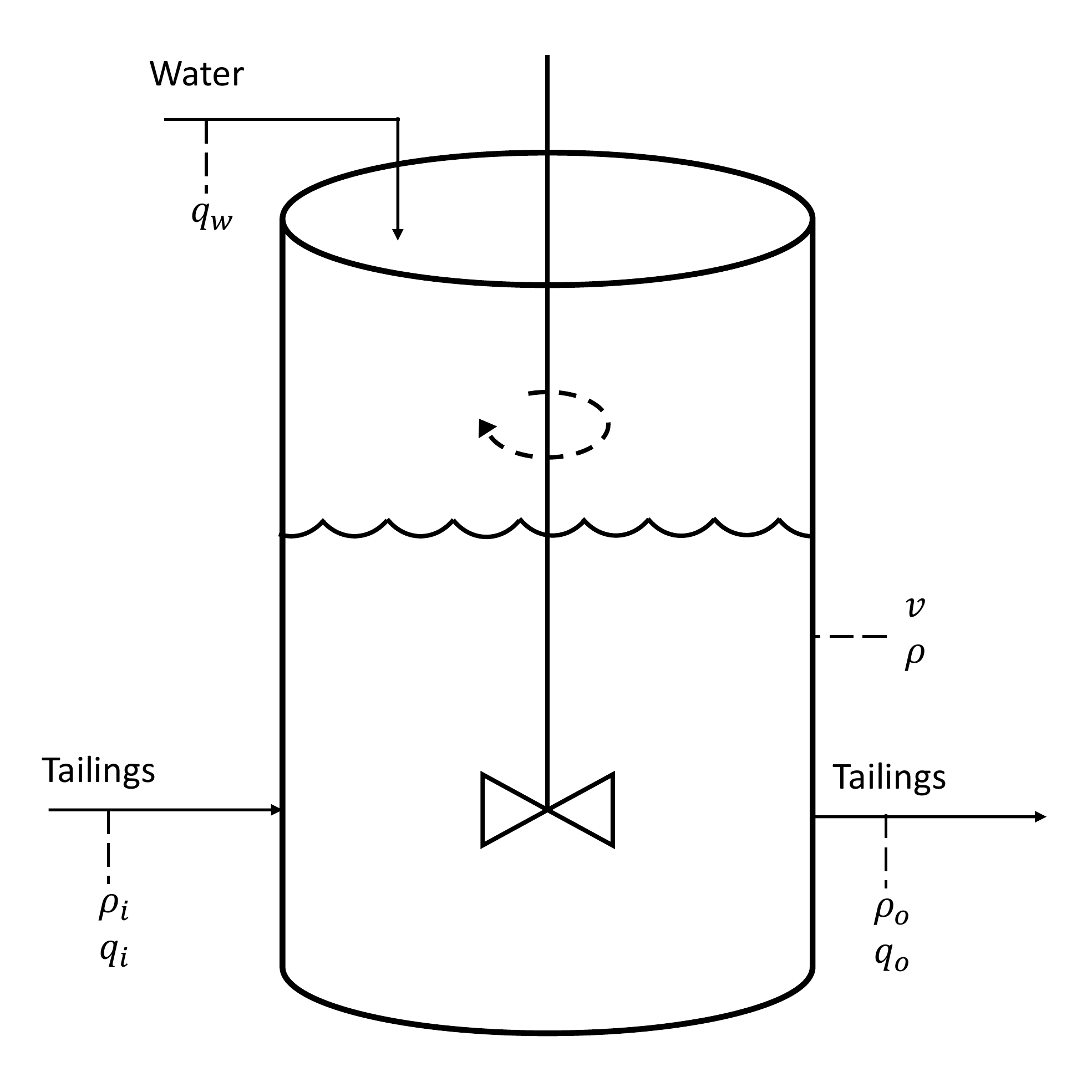}
	\caption{Bulk Tailings Treatment (BTT) surge tank.}
	\label{fig:BTTSurgeTank}
\end{figure}

The BTT surge tank is presented in Fig. \ref{fig:BTTSurgeTank}, along with all its relevant process measurements. The input flow rate, the water flow rate, and the output flow rate are respectively $q_i$, $q_w$, and $q_o$. The tank volume, which is calculated from a tank level measurement, is $v$, and the input density, the density in the tank, and the output density are $\rho_i$, $\rho$, and $\rho_o$ respectively.

The dynamic model for the surge tank is derived using a mass balance as
\begin{equation}
	\frac{d\rho v}{dt} = \rho_iq_i + q_w - \rho_o q_o.
\label{eq:MassBalance}
\end{equation}

Assuming perfect mixing, i.e. $\rho = \rho_o$, and by using the chain rule (\ref{eq:MassBalance}) can be simplified by expanding the accumulation term on the left as
\begin{equation}
	v\frac{d\rho}{dt} + \rho\frac{dv}{dt}= \rho_iq_i + q_w - \rho q_o.
\label{eq:AxpandedMassBalance}
\end{equation}

Further assuming no volume change during mixing \citep{Dontsov2014}, the volume in the surge tank will be conserved 
\begin{equation}
	\frac{dv}{dt} = q_i + q_w - q_o.
\label{eq:VolumeDynamics}
\end{equation}

Substituting (\ref{eq:VolumeDynamics}) into (\ref{eq:AxpandedMassBalance}) gives
\begin{equation}
		v\frac{d\rho}{dt}=\rho_iq_i+q_w-\rho(q_i+q_w).				 
\label{eq:NonLinearDensityDynamics}
\end{equation}
Equations (\ref{eq:VolumeDynamics}) and (\ref{eq:NonLinearDensityDynamics}) combined give the nonlinear state-space model:
\begin{equation}
\label{eq:non-lin model}
\begin{bmatrix}
\dot{v}\\\dot{\rho}
\end{bmatrix} = 
\begin{bmatrix}
q_i+ q_w- q_0 \\
\dfrac{1}{v}(\rho_iq_i+q_w-\rho(q_i+q_w))
\end{bmatrix}
\end{equation}
which is of the form:
\begin{equation}
\label{eq:non-lin ss}
\mathbf{\dot{x}}=\mathbf{f}(\mathbf{x},\mathbf{u},\mathbf{d})
\end{equation}
In this case the elements of the state vector $\mathbf{x}$ are $\rho$ and $v$, the inputs $\mathbf{u}$ are $q_i$, $q_w$ and $q_o$ and the disturbance $\mathbf{d}$ is $\rho_i$. 

Linearising (\ref{eq:non-lin model}) using a first order Taylor approximation around the equilibrium values $\mathbf{x}^*$, $\mathbf{u}^*$, $\mathbf{d}^*$ such that:
\begin{equation}
\label{eq:equilibrium}
\mathbf{f}(\mathbf{x}^*,\mathbf{u}^*,\mathbf{d}^*)=0
\end{equation}
gives:
\begin{equation}
\label{eq:lin_model}
\begin{split}
\begin{bmatrix}
\delta \dot{v}\\  \delta \dot{\rho}
\end{bmatrix} &= 
\begin{bmatrix}
0&&0 \\
-\dfrac{1}{v^2}(\rho_iq_i+q_w-\rho(q_i+q_w)) && -\dfrac{1}{v}(q_i+q_w)
\end{bmatrix}^*
\begin{bmatrix}
\delta v \\ \delta \rho
\end{bmatrix}\\ &+
\begin{bmatrix}
1&&1&&-1 \\
\dfrac{1}{v}(\rho_i-\rho) && \dfrac{1}{v}(1-\rho) && 0
\end{bmatrix}^*
\begin{bmatrix}
\delta q_i \\ \delta q_w \\ \delta q_0
\end{bmatrix} + 
\begin{bmatrix}
0 \\ \dfrac{q_i}{v}
\end{bmatrix}^* \delta p_i
\end{split}
\end{equation}

From historical data the nominal input flow rate $q_i^*$ and input density $\rho_i^*$ were determined to be $600$ $m^3/h$ and $1.5$ $t/m^3$ respectively. A target output density of $\rho = 1.4$ $t/m^3$ is chosen. This is lower than $\rho_i$ and would allow for the water and product flow rates $q_w$ and $q_i$ to be used in order to compensate for input densities below the nominal $\rho_i^*$. The tank volume $v^*$ is $10$ $m^3$ and the output flowrate $q_o$ is held constant at $750$ $m^3/h$ (i.e. $\delta q_0=0$). Substitution of these values into (\ref{eq:non-lin model}) confirms that these conditions constitute an equilibrium and (\ref{eq:equilibrium}) is satisfied. Substitution of these values into (\ref{eq:lin_model}) yields a general linear, time invariant (LTI) state-space model of the form:
\begin{equation}
\label{eq:lin_ss}
	\dot{\mathbf{x}} = \mathbf{A}\mathbf{x} + \mathbf{B}\mathbf{u} + \mathbf{G}_d \mathbf{d}
\end{equation}
\begin{equation}
	\mathbf{y} = \mathbf{C}\mathbf{x}
\end{equation}

where
\begin{equation}
\label{eq:lin_model_ss}
\begin{split}
\begin{bmatrix}
\dot{x}_1\\  \dot{x}_2
\end{bmatrix} = 
\begin{bmatrix}
0&&0 \\
0 && -75
\end{bmatrix}
\begin{bmatrix}
x_1 \\ x_2
\end{bmatrix}+
\begin{bmatrix}
1&&1 \\
0.01 && -0.04
\end{bmatrix}
\begin{bmatrix}
u_1 \\ u_2 
\end{bmatrix} + 
\begin{bmatrix}
0 \\ 60
\end{bmatrix} d
\end{split}
\end{equation}
and
\begin{equation}
\begin{bmatrix}
y_1\\  y_2
	\end{bmatrix} = 
	\begin{bmatrix}
	1&&0 \\
	0 && 1
	\end{bmatrix}
	\begin{bmatrix}
	x_1 \\ x_2
	\end{bmatrix}
\end{equation}

Converting (\ref{eq:lin_model_ss}) to a transfer function matrix model of the form:
\begin{equation}
 \mathbf{y} = 
\mathbf{G}_p(s)
\mathbf{u}+
\mathbf{G}_d(s) d 
\end{equation}
yields:
\begin{equation}
\label{eq:lin_model_tf}
\begin{split}
\\
\begin{bmatrix}
y_1\\  y_2
\end{bmatrix} 
 = 
\begin{bmatrix}
\dfrac{1}{s}&&\dfrac{1}{s}\\[12pt]
\dfrac{0.01}{s+75} && \dfrac{-0.04}{s+75}
\end{bmatrix}
\begin{bmatrix}
u_1 \\ u_2
\end{bmatrix}+
\begin{bmatrix}
0 \\[6pt] \dfrac{60}{s+75}
\end{bmatrix} d
\end{split}
\end{equation}

where the outputs $y_1$ and $y_2$ are equivalent to the states $x_1$ and $x_2$ and thus correspond directly with $v$ and $\rho$ respectively. 

Table \ref{tab: tank par} shows a summary of the nominal parameter values as well as the corresponding maximum and minimum values.
\renewcommand{\arraystretch}{1.3}
\begin{table}[h]
	\centering
	\caption{Parameter constraints and nominal values}
	\label{tab: tank par}
	\begin{tabular}{c c c c c}
		\hline
		\textbf{Variable}&Nominal&Minimum&Maximum&Unit\\ 
		$v$ & 10 & 3 & 20 & $m^3$\\ 
		$\rho$ & 1.4 & 1 & 1.5 & $t/m^3$\\
		$q_i$ & 600 & 300 & 1200 & $m^3/h$\\
		$q_w$ & 150 & 0 & 750 & $m^3/h$\\
		$\rho_i$ & 1.5 & 1 & 2 & $t/m^3$\\
		\hline
	\end{tabular}
\end{table}
\renewcommand{\arraystretch}{1}

An input/output controllability analysis in presented in Appendix \ref{Appendix: IO cntrl} where it is shown that it is possible to keep the plant outputs $\mathbf{y}$ within the specified bounds from the reference $\mathbf{r}$ using the available actuator authority $\mathbf{u}$ in the presence of disturbances $\mathbf{d}$. It is also shown that the reference density will only become unachievable if the input density $\rho_i$ falls below this reference value. In terms of input output pairing for decentralised control strategies, it is also shown from the relative gain array in (\ref{eq:RGA_plant}) that it would be best to use the product flow rate $q_i$ to control the tank volume $v$ and the flow rate of water into the tank $q_w$ to control the density of the liquid in the tank $\rho$.  

\subsection{The Vendors} \label{Section:Controllers}

Three different vendors act in competition for control of the plant in this case study. Each vendor employs a different APC strategy with one of the three using conventional feedback control ($\mathbf{G}_{c1}$), namely an inverse based controller, and the remaining two use MPC controllers, one linear ($\mathbf{G}_{c2}$) and the other nonlinear ($\mathbf{G}_{c3}$). A decentralised PI controller as would be used in practice is employed as the local fallback controller ($\mathbf{G}_{c\_local}$) as depicted in Fig.~\ref{fig:Competing_APC_2}. 

As mentioned in Section \ref{subsec: stability}, it is required to have some minimum bandwidth in order to ensure system stability. All the controllers were thus designed to yield as similar a response as possible with a desired bandwidth of 100 $rad/s$ being used in the synthesis of the conventional feedback controllers. It should be mentioned that this minimum bandwidth would be specified either exclusively by the plant or by the plant in conjunction with the platform service provider.  

Each controller used by the three different vendors as well as the local fallback controller are presented in the following subsections. 

\subsubsection{$\mathbf{G}_{c\_local}$ -- De-coupled PI Controller}\

\begin{figure}[h!]
	\centering
	\includegraphics[width=9.5 cm]{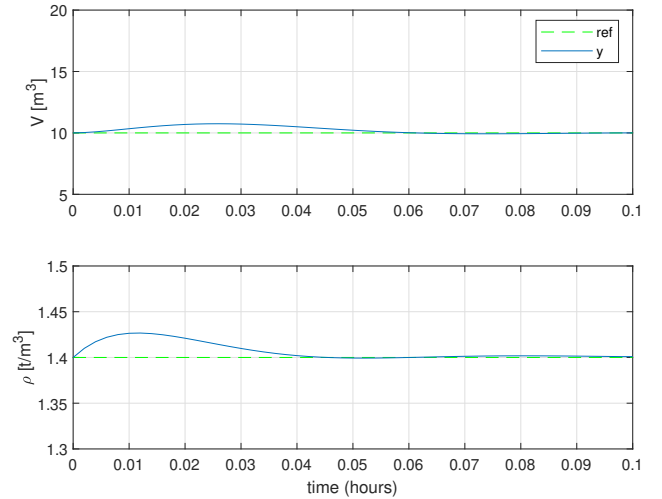}
	\caption{Plant outputs resulting from the de-coupled PI controller ($\mathbf{G}_{c\_local}$) for a step disturbance of 0.1 $t/m^3$ in $\rho_i$ with 10\% gain uncertainty in $q_w$.}
	\label{fig:tank_y_PI}
\end{figure}

A typical control solution which may be applied in practice is to use de-coupled PI controllers which are tuned according to some rules such as those given in Table \ref{tab: PI rules}.
\renewcommand{\arraystretch}{2.2}
\begin{table}[h]
	\centering
	\caption{PI Tuning Rules}
	\label{tab: PI rules}
	\begin{tabular}{c c c c}
		\hline
		\textbf{Model}&\textbf{G}(s)&$K_c$&$\tau_i$\\ 
		First order & $k\dfrac{e^{-\theta s}}{\tau s+1}$ & $\dfrac{3\tau}{kT_R}  $ & $\tau$	\\ 
		Integrator & $k\dfrac{e^{-\theta s}}{s}$ & $\dfrac{4.2}{kT_R}  $ & $0.4T_R$ 	\\
		\hline
	\end{tabular}
\end{table} 
According to the relative gain array given in (\ref{eq:RGA_plant}), it is best to pair the inputs in a diagonal fashion with the input flow rate $q_i$ used to control the tank volume $v$ and the water flow rate $q_w$ used to control the liquid density $\rho$. The tuning factor $T_R$ in Table \ref{tab: PI rules} serves as a tuning parameter to adjust the aggressiveness of the controller which was chosen to be 0.05. Applying these rules to the relevant models ($\mathbf{G}_p(1,1)$ and $\mathbf{G}_p(2,2)$ in (\ref{eq:lin_model_tf})) for each input-output pairing, the following controller is derived:
\renewcommand{\arraystretch}{1}
\begin{equation}
\label{eq:controller_3}
\mathbf{G}_{c\_local} = 
\begin{bmatrix}
\dfrac{84(s+50)}{s}&&0\\[12pt]
0 && \dfrac{-1505.7(s+75)}{s}
\end{bmatrix}
\end{equation}
Fig. \ref{fig:tank_y_PI} shows the response of the de-coupled PI controller ($\mathbf{G}_{c\_local}$) for a step disturbance of 0.1 $t/m^3$ in $\rho_i$ with 10\% gain uncertainty in $q_w$. It should be noted that the controller $\mathbf{G}_{c\_local}$ is simulated in closed-loop using the \emph{full nonlinear plant model} of (\ref{eq:non-lin model}) and this is also true for all subsequent simulations. It is evident from these figures that the controller adequately rejects the disturbance, bringing both the volume and density back to set-point within 0.06 hours.  

%\begin{figure}[h!]
%	\centering
%	\includegraphics[width=9.5 cm]{tank_u_PI.eps}
%	\caption{Plant inputs resulting from the de-coupled PI controller ($\mathbf{G}_{c\_local}$) for a step disturbance of 0.1 $t/m^3$ in $\rho_i$ with 10\% gain uncertainty in $q_w$.}
%	\label{fig:tank_u_PI}
%\end{figure} 

\subsubsection{$\mathbf{G}_{c1}$ -- Modified Inverse Based Controller}\

Ideal control is defined by the ability to incorporate the inverse of the plant model in the controller and by this means achieve the loop transfer function:
\begin{equation}
\label{eq:loop_tf}
\mathbf{L} = \mathbf{G}_{c} \mathbf{G}_p = \frac{k}{s}\mathbf{I}
\end{equation}
where $k$ is an adjustable gain parameter and $\mathbf{I}$ is the identity matrix. For this to be achieved the controller has to take on the form:
\begin{equation}
\mathbf{G}_{c}=\frac{k}{s}\mathbf{G}_p^{-1}
\end{equation}
This controller is chosen as it is perhaps the simplest and cheapest APC possible, so it might be favoured by the selector based on price in the case that the selector performance index contains a price component. 

Ideal control is only achievable if the plant model $\mathbf{G}_p$ is square, does not contain any time delays or right half-plane zeros, or if the plant denominator polynomials are of a degree two or more higher than the numerator polynomials \citep{Skogestad2005}. None of these impediments are present in the plant model given in (\ref{eq:lin_model_tf}) and the controller is determined as:
\begin{equation}
\label{eq:controller_1}
\mathbf{G}_{c} = k
\begin{bmatrix}
0.8&&\dfrac{20(s+75)}{s}\\[12pt]
0.2 && \dfrac{-20(s+75)}{s}
\end{bmatrix}
\end{equation}
The gain factor $k$ is used to adjust the bandwidth of the controller. It can be seen that the resulting controller consists of two purely proportional elements and two elements which take on a PI structure, with the proportional elements acting only on the volume error and the PI elements acting only on the density error. 

The notion of ideal control is evidently only valid if the plant is purely linear and there is no uncertainty in the model, which are both impractical assumptions. It is found however for the surge tank under consideration, that the inverse controller in (\ref{eq:controller_1}) yields satisfactory performance when simulated on the nonlinear model. There is however an issue encountered in the case where there is any gain uncertainty on the two input channels $q_i$ and $q_w$. This results in a steady state error in the tank volume $v$ in response to either disturbances from $\rho_i$ or set-point changes in the output variables $v$ or $\rho$.

\begin{figure}[h!]
	\centering
	\includegraphics[width=9.5 cm]{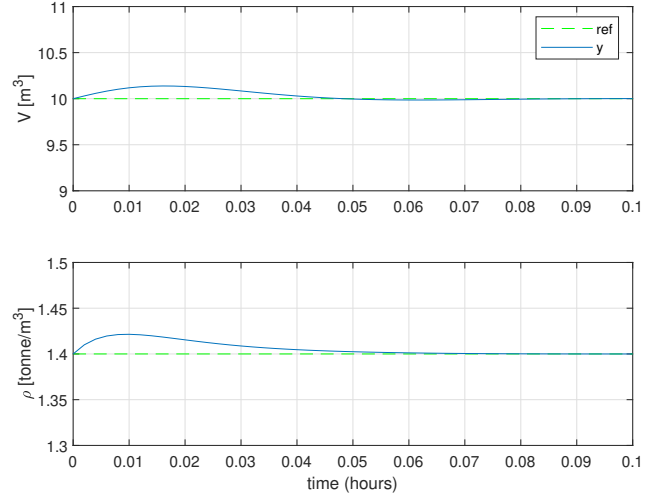}
	\caption{Plant outputs resulting from the modified inverse controller ($\mathbf{G}_{c1}$) for a step disturbance of 0.1 $t/m^3$ in $\rho_i$ with 10\% gain uncertainty in $q_w$.}
	\label{fig:tank_y_mod_inv}
\end{figure}

The steady-state error phenomenon is due to the purely proportional terms which act on the volume error. This can easily be overcome by the addition of integral action to these two terms. The controller then takes the from:
\begin{equation}
\label{eq:controller_2}
\mathbf{G}_{c1} = 100
\begin{bmatrix}
\dfrac{0.8(s+41.3)}{s}&&\dfrac{20(s+75)}{s}\\[12pt]
\dfrac{0.2(s+165)}{s} && \dfrac{-20(s+75)}{s}
\end{bmatrix}
\end{equation}
where the gain of the additional integral terms was chosen to be 33.  Fig. \ref{fig:tank_y_mod_inv} shows the response of this controller to an increase of 0.1 $t/m^3$ in $\rho_i$ with the same 10\% gain uncertainty in $q_w$. It can be seen that the steady state error in the volume $v$ is eliminated by the additional integral action.  
%The corresponding plant inputs are shown in Fig. \ref{fig:tank_u_mod_inv}. 

%\begin{figure}[h!]
%	\centering
%	\includegraphics[width=9.5 cm]{mod_inv_inputs.pdf}
%	\caption{Plant inputs resulting from the modified inverse controller ($\mathbf{G}_{c1}$) for a step disturbance of 0.1 $t/m^3$ in $\rho_i$ with 10\% gain uncertainty in $q_w$.}
%	\label{fig:tank_u_mod_inv}
%\end{figure} 

It can also be seen from Figs. \ref{fig:tank_y_PI} and \ref{fig:tank_y_mod_inv} that the response in the density $\rho$ resulting from the use of the decentralised PI controller $\mathbf{G}_{c\_local}$ is comparable with that of controller $\mathbf{G}_{c1}$.

\subsubsection{$\mathbf{G}_{c2}$, $\mathbf{G}_{c3}$ -- Model Predictive Control} \label{sec: mpc}\

MPC is an advanced control technique which has been put to use extensively in the process control industry (\cite{Forbes2015}; \cite{Qin2003}), and has also received much attention from the academic community (\cite{Camacho1999}; \cite{Mayne2009}). The concept of MPC involves the optimisation of a performance function subject to the plant dynamics as well as constraints on the input and output variables over some finite prediction horizon. This process is repeated at each sampling period. This can be expressed in mathematical terms as follows.
\begin{displaymath}	
\underset{\mathbf{u},\delta}{min}\quad V(\mathbf{x}, \mathbf{u},\boldsymbol{\delta},\mathbf{y}_{ref}(k))
\end{displaymath}
subject to
\begin{displaymath}
\mathbf{x}(j+k+1|k)  = \mathbf{f}(\mathbf{x}(j+k|k),\mathbf{u}(i+k|k))
\end{displaymath}
\begin{displaymath}
\mathbf{y}(j+k|k)  = \mathbf{g}(\mathbf{x}(j+k|k),\mathbf{u}(i+k|k))
\end{displaymath}
\begin{displaymath}
\mathbf{x}(k|k) = \mathbf{x}_0
\end{displaymath}
\begin{displaymath}
\mathbf{u}_{min}\le \mathbf{u}(i+k|k) \le \mathbf{u}_{max} 
\end{displaymath}
\begin{displaymath}
\mathbf{y}_{min}-\boldsymbol{\delta}(k) \le \mathbf{y}(j+k|k) \le \mathbf{y}_{max}+\boldsymbol{\delta}(k) 
\end{displaymath}
\begin{displaymath}
k = \textrm{Present sampling instant.} 
\end{displaymath}
\begin{displaymath}
j = 0,1,..., N_p-1 
\end{displaymath}
\begin{displaymath}
i = 0,1,..., N_c-1
\end{displaymath}
\begin{displaymath}
N_p \ge N_c \times N_b
\end{displaymath}
\begin{displaymath}
\mathbf{x}\in \Re^{N_x \times N_p}
\end{displaymath}
\begin{displaymath}
\mathbf{u}\in \Re^{N_u \times N_c}
\end{displaymath}
\begin{displaymath}
\mathbf{y}\in \Re^{N_y \times N_p}
\end{displaymath}
\begin{displaymath}
\boldsymbol{\delta}\in \Re^{N_y}
\end{displaymath}
where $N_x$ is the number or states, $N_u$ is the number of control inputs, $N_y$ is the number of outputs, $N_p$ is the number of predictions in the prediction horizon, $N_c$ is the number of control moves to be calculated in the optimisation problem, $N_b$ is the number of subsequent samples the control move is kept constant during predictions and $\mathbf{x}_0$ is the value of the state at the present sampling time. The performance index $V$ typically takes on a quadratic form such as:
\begin{equation}
\label{eq:MPC problem}
\begin{split}
V(\mathbf{u},\mathbf{x},\boldsymbol{\delta},\mathbf{y}_{ref}(k))&=\\
\sum_{j=1}^{N_p}(\mathbf{y}_{ref}(k)-\hat{\mathbf{y}}(k+j|k))^{T}&\mathbf{Q}(\mathbf{y}_{ref}(k)-\hat{\mathbf{y}}(k+j|k))\\+\sum_{i=1}^{N_c}\Delta \mathbf{u}(i+k|k)^{T}\mathbf{R}\Delta \mathbf{u}&(i+k|k)
+\boldsymbol{\delta}(k)^T \boldsymbol{\Psi} \boldsymbol{\delta}(k)\\
\end{split}
\end{equation}
where matrices $\mathbf{Q}$, $\mathbf{R}$ and $\boldsymbol{\Psi}$ are diagonal matrices used to weight the importance of the different input, output and slack variables which in turn adjusts the behaviour of the controller. $\hat{\mathbf{y}}$ indicates the predicted values of the process output and: 
\begin{displaymath}
\mathbf{y}_{ref} \in \Re^{N_y}
\end{displaymath}
\begin{displaymath}
\mathbf{Q}\in \Re^{N_y \times N_y}
\end{displaymath}
\begin{displaymath}
\mathbf{R} \in \Re^{N_u \times N_u}
\end{displaymath}
\begin{displaymath}
\boldsymbol{\Psi}\in \Re^{N_y \times N_y}
\end{displaymath}
\begin{displaymath}
\Delta \mathbf{u}(i) = \mathbf{u}(i+k|k) - \mathbf{u}(i+k-1|k)
\end{displaymath}

At each sampling instant a measurement of the system state $\mathbf{x}_0$ is taken. The measured state is then used as an initial condition for model predictions to be performed as part of the optimisation procedure. The specified number $N_c$ of control vectors as well as the vector of slack variables $\boldsymbol{\delta}$ are the decision variables in the optimisation procedure. The input column vectors of $\mathbf{u}$ are used to simulate the plant model (either nonlinear or linear) with $N_b$ denoting the number of subsequent time steps the input vector is kept constant. The vector of slack variables $\boldsymbol{\delta}$ is used as a decision variable in the optimisation with the purpose of implementing soft constraints on the output variables. This allows the output to violate the constraints at times with the slack variables used to drive them back slowly to within the constraints. Hence the elements in the weighting matrix $\boldsymbol{\Psi}$ are typically large. The solution produces $N_c$ control vectors which can be denoted as $\tilde{\mathbf{u}}$. The first vector in $\tilde{\mathbf{u}}$, that is $\tilde{\mathbf{u}}(0)$ is implemented on the plant at the beginning of the next time step, when a new measurement of the state is taken and the optimisation process is repeated.

It is often the case that not all the states in $\mathbf{x}$ are measurable, which has given rise to the necessity of state estimation by the use of some form of state observer. In this case it is not necessary as the states in both the linear and nonlinear state-space models in (\ref{eq:non-lin model}) and (\ref{eq:lin_model}) are the same as the outputs of the system and are measurable. 

The MPC problem formulation described will provide off-set free reference tracking, provided that the model is an ideal representation of the plant \citep{Mayne2009}. If there is any plant model mismatch or unmeasured disturbances, steady state off-set will result due to differences in the model predictions and actual plant behaviour. The input density $\rho_i$ is not available to the controllers, and is thus an unmeasured disturbance. In order to correct for this, some type of integral mechanism needs to be included in the controller. 

The most commonly used method to achieve zero steady state off-set is output bias correction (\cite{Camacho1999}; \cite{Meadows}). The method used in this work however is to create additional states which serve as input disturbance estimates for each input and make use of an observer to estimate these states as described in \cite{Mayne2009}. A linear observer is used to perform the state estimation which makes use of the augmented discrete linear system given as:
\begin{equation}
\label{eq:ss_aug}
\begin{bmatrix}
\mathbf{x}(k+1) \\ \mathbf{v}(k+1) 
\end{bmatrix}
=
\begin{bmatrix}
\boldsymbol{\Phi} && \boldsymbol{\Gamma}  \\
\mathbf{0} && \mathbf{I} \\
\end{bmatrix}
\begin{bmatrix}
\mathbf{x}(k) \\ \mathbf{v}(k) 
\end{bmatrix} + 
\begin{bmatrix}
\boldsymbol{\Gamma}\\ \mathbf{0} 
\end{bmatrix} \mathbf{u}(k) + \begin{bmatrix}
\mathbf{0} \\ \mathbf{I}
\end{bmatrix}\mathbf{w}(k)
\end{equation}
\begin{equation}
\label{eq:ss_aug2}
\mathbf{y}(k) = 
\begin{bmatrix}
\mathbf{C} && \mathbf{0} 
\end{bmatrix}
\begin{bmatrix}
\mathbf{x}(k) \\ \mathbf{v}(k) 
\end{bmatrix} + \mathbf{n}(k)
\end{equation}
where $\mathbf{v} \in \Re^{N_u \times 1} $ is the vector of input disturbance estimates, $\mathbf{w}$ and $\mathbf{n}$ are white noise for the purpose of observer design and $\boldsymbol{\Phi}$ and $\boldsymbol{\Gamma}$ are the discrete state-space coefficient matrices determined from (\ref{eq:lin_ss}) according to (\ref{eq:Phi-discrt}) and (\ref{eq:Gam-discrt}). The augmented system in (\ref{eq:ss_aug}) and (\ref{eq:ss_aug2}) is rewritten as:
\begin{equation}
\label{eq:ss_aug3}
\tilde{\mathbf{x}}(k+1) = \tilde{\boldsymbol{\Phi}} \tilde{\mathbf{x}}(k) + \tilde{\boldsymbol{\Gamma}}\mathbf{u}(k) + \mathbf{H}\mathbf{w}(k)
\end{equation}
\begin{equation}
\label{eq:ss_aug4}
\mathbf{y}(k) = 
\tilde{\mathbf{C} }
\tilde{\mathbf{x}}(k) + \mathbf{n}(k)
\end{equation}
where $\tilde{\mathbf{x}}$ is the augmented state consisting of the disturbance estimates $\mathbf{v}(k)$. The state estimation is then performed as follows:
\begin{equation}
\label{eq:observer1}
\hat{\tilde{\mathbf{x}}}(k) = \hat{\tilde{\mathbf{x}}}_*(k) + \mathbf{L}(\mathbf{y}_m(k)-\tilde{\mathbf{C}}\hat{\tilde{\mathbf{x}}}(k))
\end{equation}
\begin{equation}
\label{eq:observer2}
\hat{\tilde{\mathbf{x}}}_*(k+1) = \tilde{\boldsymbol{\Phi}} \hat{\tilde{\mathbf{x}}}(k) + \tilde{\boldsymbol{\Gamma}}\mathbf{u}(k) 
\end{equation}
where $\hat{\tilde{\mathbf{x}}}_*$ and $\hat{\tilde{\mathbf{x}}}$ are the predicted and updated estimates of the augmented state $\tilde{\mathbf{x}}$ and $\mathbf{L}$ is the observer gain \citep{Mayne2009}. The estimates of the fictitious input disturbance at each time step denoted as $\hat{\mathbf{v}}(k)$ are then added directly to the inputs during the model prediction portion of the MPC optimisation algorithm. This effectively yields integral action in a similar manner as has been shown for state feedback control \citep{Astrom1997}. The observer gain $\mathbf{L}$ was determined through the use of the Kalman filtering technique which, due to the augmentation of the system in (\ref{eq:ss_aug}) allows for the aggressiveness of the integral action to be manipulated by the covariance matrices of the process and measurement noise denoted as $\mathbf{Q}_w$ and $\mathbf{R}_n$ respectively. 

Two MPC controllers were designed with one using the linear plant model to perform predictions ($\mathbf{G}_{c2}$) and the other using the nonlinear model ($\mathbf{G}_{c3}$). For both controllers, the weighting matrices $\mathbf{Q}$ and $\mathbf{R}$ used to tune the controller and the slack variable weighting matrix $\boldsymbol{\Psi}$ were chosen to be:
\begin{displaymath}
\mathbf{Q} = \begin{bmatrix}
1e-3&&0 \\ 0&&1
\end{bmatrix}, \quad \mathbf{R} = 0.5e-7 \times \mathbf{I}, \quad \boldsymbol{\Psi} = 1e7 \times \mathbf{I}
\end{displaymath}
The error of the tank volume $v$ thus contributes much less to the value of the objective function than the density $\rho$, with changes in the control action contributing the least in order to allow for more aggressive control action. The covariance matrices $\mathbf{Q}_w$ and $\mathbf{R}_n$ were chosen to be:
\begin{displaymath}
\mathbf{Q}_w = \mathbf{I}, \quad \mathbf{R}_n =1e-5\times\mathbf{I}
\end{displaymath}
The larger the ratio between $\mathbf{Q}_w$ and $\mathbf{R}_n$, the more aggressive the observer and the quicker it will react to disturbances, thus the filter is tuned to be relatively aggressive. 

\begin{figure}[h!]
	\centering
	\includegraphics[width=9.5 cm]{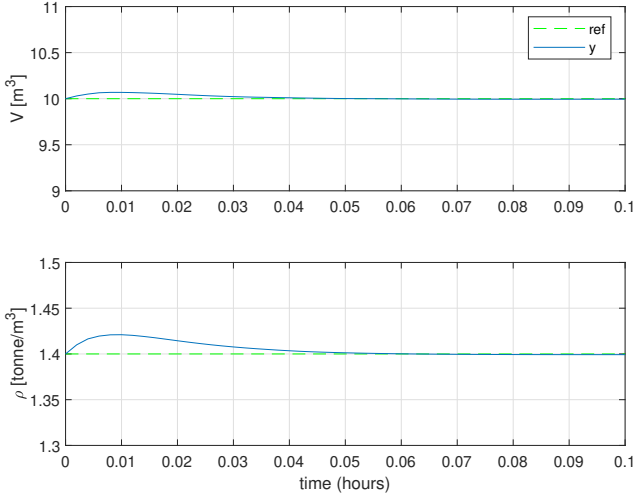}
	\caption{Plant outputs resulting from the nonlinear MPC controller ($\mathbf{G}_{c3}$) for a step disturbance of 0.1 $t/m^3$ in $\rho_i$ with 10\% gain uncertainty in $q_w$.}
	\label{fig:tank_y_nmpc}
\end{figure}

\begin{figure}[h!]
	\centering
	\includegraphics[width=9.5 cm]{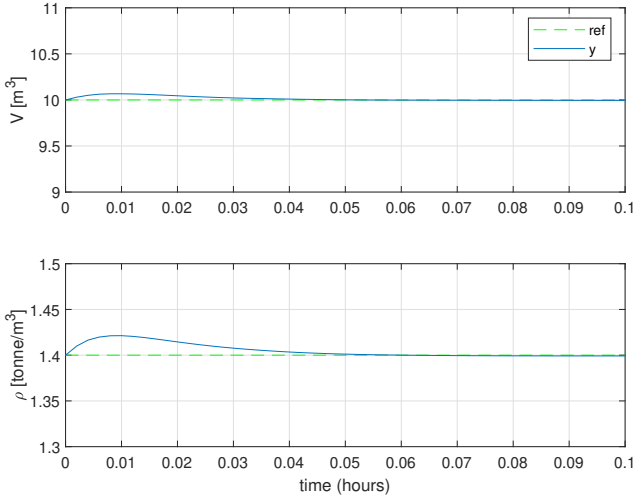}
	\caption{Plant outputs resulting from the linear MPC controller ($\mathbf{G}_{c2}$) for a step disturbance of 0.1 $t/m^3$ in $\rho_i$ with 10\% gain uncertainty in $q_w$.}
	\label{fig:tank_y_lmpc}
\end{figure}

Figs. \ref{fig:tank_y_nmpc} and \ref{fig:tank_y_lmpc} show the response of the nonlinear and linear MPC controllers respectively to an increase of 0.1 $t/m^3$ in $\rho_i$ with 10\% gain uncertainty in $q_w$. 

It can be seen from these figures that the response is practically identical. The reason for this can be attributed to the fact that the observer is predominantly responsible for the rejection of disturbances as well as the fact that the linear model is very accurate close to the operating point. The tuning parameters $\mathbf{Q}$, $\mathbf{R}$, $\mathbf{Q}_w$ and $\mathbf{R}_n$ were explicitly chosen to yield a response as similar as possible to that achieved by $\mathbf{G}_{c1}$, particularly for the density $\rho$.

\subsection{Facilitating Platform} \label{Section:Competing}
The platform which facilitates the competition of the three APCs presented in the previous section to control the process in the manner depicted in Fig.~\ref{fig:Competing_APC_2}, is presented in this section. It is assumed that the disturbance $d$ (which is the density of the feed $\rho_i$ into the tank) is measured, and therefore known to the selector, but is not made available to the vendors. 

\begin{figure}[h!]
	\centering
	\includegraphics[width=9.5 cm]{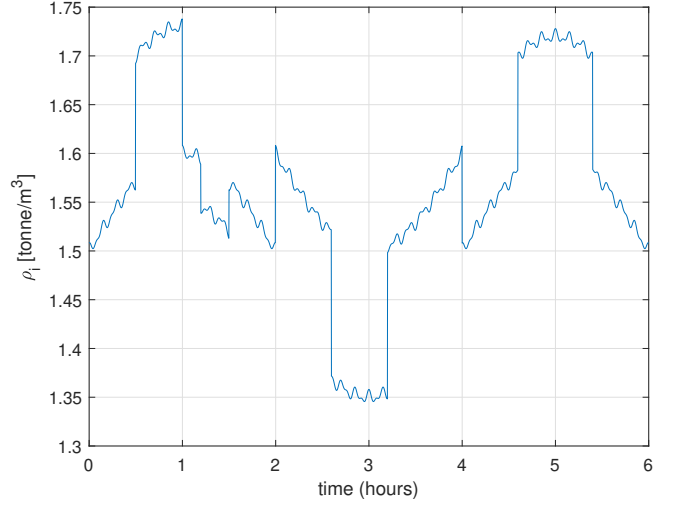}
	\caption{Disturbance variable $\rho_i$.}
	\label{fig:tank_comp_d}
\end{figure}

The local fallback controller $\mathbf{G}_{c\_local}$ is taken to be the de-coupled PI controller as this is the simplest and most generic control structure which would be employed by plant personnel in practice. All other controllers are developed by independent vendors and are assumed to reside in a cloud platform that could be in a remote location. Fig. \ref{fig:tank_comp_d} shows the value of the disturbance variable $\rho_i$ over the simulation period of 6 hours which takes on a maximum value of 1.74 $t/m^3$ and a minimum of 1.35 $t/m^3$. It can be seen that for a brief period the density falls below the set-point of 1.4 $t/m^3$ for which density setpoint following is unachievable.

The nomenclature of the controllers used is:
\begin{enumerate}[start=0]
	\item De-coupled PI controller ($\mathbf{G}_{c\_local}$)
	\item Modified inverse controller ($\mathbf{G}_{c1}$)
	\item Linear MPC controller ($\mathbf{G}_{c2}$)
	\item nonlinear MPC controller ($\mathbf{G}_{c3}$)
\end{enumerate}
The selector in Fig.~\ref{fig:Competing_APC_2} determines the best controller in the manner described in Section \ref{sec: performance measure}, using the nonlinear plant model as the plant model to simulate the closed loop performance of the controllers, together with the performance index in (\ref{eq:min sse}) for evaluation. The weighting matrix $\mathbf{W}_e$ in (\ref{eq:min sse}) was set to:
\begin{equation}
\mathbf{W}_e = \mathbf{Q} = \begin{bmatrix}
1e-3&&0 \\ 0&&1
\end{bmatrix}
\end{equation}  
which is the same weighting matrix $\mathbf{Q}$ used in the performance index of the two MPC controllers, placing significantly more emphasis on the density $\rho$ than the volume $v$. It is thus assumed that the performance measure in (\ref{eq:min sse})  and the weighting matrix $\mathbf{W}_e$ is shared with the controllers, but this does not necessarily have to be the case. The evaluation horizon $N_e$ was set to 0.5 hours in order to be several orders of magnitude greater than that corresponding to the specified bandwidth of the controllers. This ensures the evaluation horizon is much longer than the expected dynamics of the slowest closed loop system.

A sampling period $\Delta t$ of 0.002 hours was chosen to satisfy the guideline that 
\begin{displaymath}
0.2<\omega_B \cdot \Delta t <0.6
\end{displaymath}  
where $\omega_B$ is the controller bandwidth (which is taken as 100 $rad/hour$) and $\Delta t$ is the sampling time of 0.002 hours \citep{Astrom1997}. The conventional feedback controllers $\mathbf{G}_{c\_local}$ and $\mathbf{G}_{c1}$ were transformed into state-space form and discretised as given in (\ref{eq:cntrl_ss_d1}) and (\ref{eq:cntrl_ss_d2}). The simulation of the nonlinear model was done using a $4^{th}$ order Runge-Kutta approximation. 

Bumpless transfer is performed by the MPCs ($\mathbf{G}_{c2}$ and $\mathbf{G}_{c3}$) and the conventional feedback controllers ($\mathbf{G}_{c\_local}$ and $\mathbf{G}_{c1}$) as described in Section \ref{sec:bumpless transfer}. This is also the case in terms of constraint handling as described in Section \ref{sec: constraint handling}. For the MPC controllers this is handled as part of the optimisation algorithm as described in Section \ref{sec: mpc}. For $\mathbf{G}_{c\_local}$ and $\mathbf{G}_{c1}$, if the controller tries to increase the flow-rate of the product feed $q_i$ above 750 $m^3/hour$, this flow-rate will be capped at this upper constraint and the water flow-rate $q_w$ will be set to 0 $m^3/hour$ in order to keep the inflow equal to the outflow. In terms of limits on the rate of change constraints on the MVs, it is assumed that both the product and water flow-rates $q_i$ and $q_w$ can be increased by a maximum  of 100 $m^3/hour$ at each sampling instant which corresponds to 100 $m^3/hour$ every 0.002 hours or 7.2 seconds (actuator and sensor dynamics are neglected).

Security and trust as described in Section \ref{subsec: security} and remuneration as described in Section \ref{subsec: renumeration} are not considered in the simulation. Rather, the purpose of this simulation is to illustrate the platform concept of providing the plant access to the best performing APC based on its performance in controlling a plant model. 

\subsection{Platform Simulation Results}

\begin{figure}[h]
	\centering
	\includegraphics[width=9.5 cm]{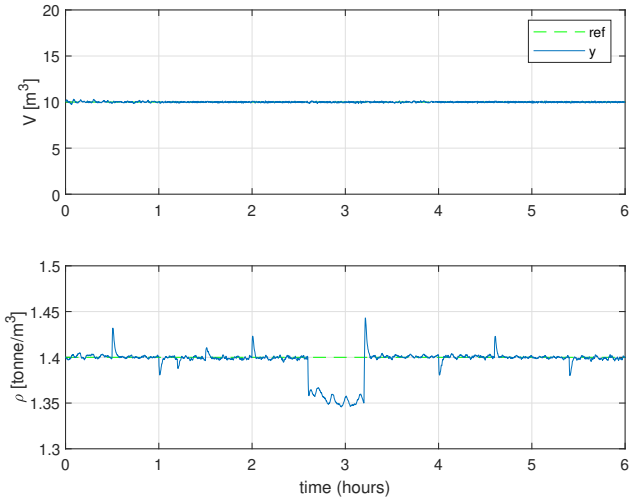}
	\caption{Plant outputs for platform control.}
	\label{fig:tank_comp_y}
\end{figure}

\begin{figure}[h]
	\centering
	\includegraphics[width=9.5 cm]{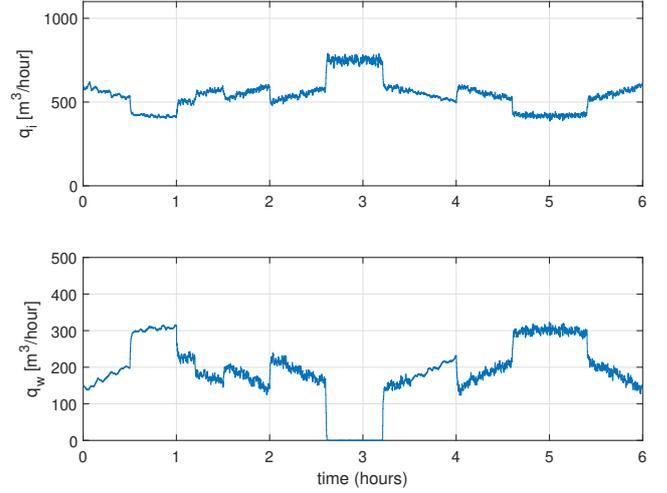}
	\caption{Plant inputs for platform control.}
	\label{fig:tank_comp_u}
\end{figure}

\begin{figure}[h]
	\centering
	\includegraphics[width=9.5 cm]{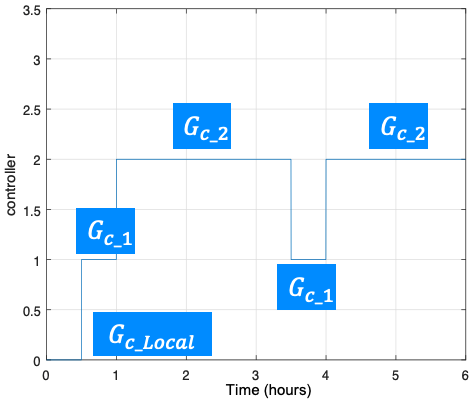}
	\caption{Selected controller.}
	\label{fig:c_2}
\end{figure}

Figs. \ref{fig:tank_comp_y} and \ref{fig:tank_comp_u} show the plant outputs and the inputs sent to the plant respectively. Fig. \ref{fig:c_2} shows the controller selected over the 6-hour period. Plant model mismatch was created by adding a 10\% gain uncertainty to the input water flow-rate $q_w$. 

As can be seen in Fig. \ref{fig:c_2}, control is initially performed by ($\mathbf{G}_{c\_local}$) (de-coupled PI), after which control is switched over to controller $\mathbf{G}_{c1}$ (modified inverse). For the remainder of the simulation period, control is performed by controller $\mathbf{G}_{c2}$ (linear MPC) apart from a single evaluation period where control is switched back over to ($\mathbf{G}_{c1}$) between the 3.5 and 4 hour mark. In the period from 2.6 hours to 3.2 hours the set-point density of 1.4 $t/m^3$ is not achievable as the input density $\rho$ decreases below 1.4 $t/m^3$ as can be seen in Fig. \ref{fig:tank_comp_d}, hence the deviation from set-point in the output $\rho$. During this period, the constraints are taken into account by the selected controller $\mathbf{G}_{c2}$ (linear MPC). The rate of change constraints on the inputs are also never violated by an active controller, with maximum changes between sampling periods of 38.7 and 46.3 $m^3/hour$ for $q_i$ and $q_w$ respectively. The spikes seen in the density $\rho$ correspond with the step changes in the input feed density $\rho_i$ as seen in Fig. \ref{fig:tank_comp_d}.

\begin{figure}[h]
	\centering
	\includegraphics[width=9.5 cm]{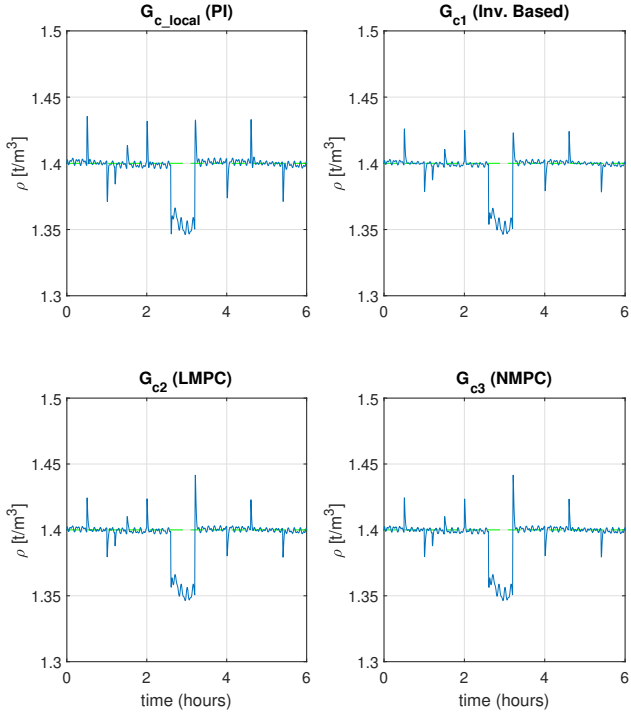}
	\caption{Plant model density $\rho$ simulated by the platform.}
	\label{fig:tank_comp_y1_sim}
\end{figure}

\begin{figure}[h]
	\centering
	\includegraphics[width=9.5 cm]{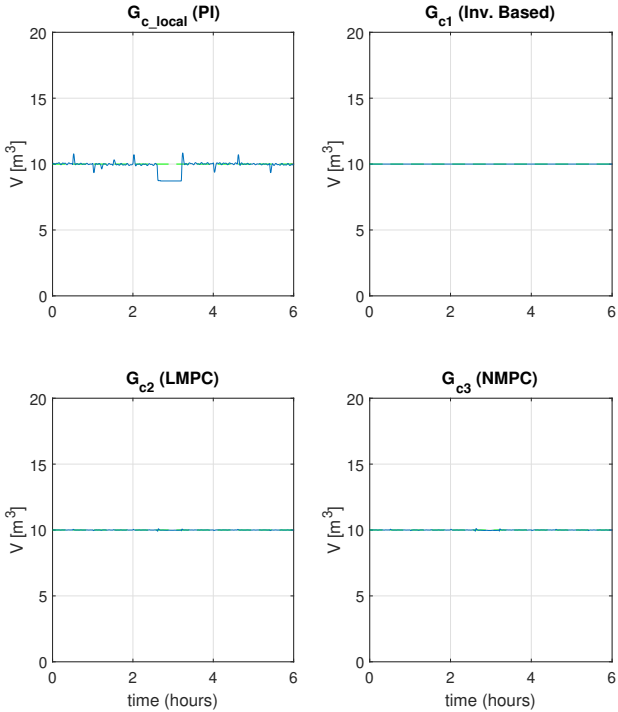}
	\caption{Plant model volume $v$ simulated by the platform.}
	\label{fig:tank_comp_y2_sim}
\end{figure}

Figs. \ref{fig:tank_comp_y1_sim} and \ref{fig:tank_comp_y2_sim} show the outputs simulated by the platform on the nonlinear plant model (no plant model mismatch) for each of the six controllers. It can be seen in Fig. \ref{fig:tank_comp_y1_sim} that the response in density $\rho$ is fairly similar for all controllers with the magnitude of the spikes in the MPC controllers slightly larger than those of the other feedback controllers. Fig. \ref{fig:tank_comp_y2_sim} shows that all controllers keep the level constant when controlling the perfect plant model except the local controller $\mathbf{G}_{c\_local}$ which performs very poorly in terms of level control. 

In order to provide a quantitative measure of overall system performance, the $SSE$ is calculated over the whole simulation according to: 
\begin{equation}
\label{eq:min sse total}
SSE_i=\sum_{k=1}^{N}(r_i(k)-y_i(k))\cdot(r_i(k)-y_i(k)),	
\end{equation}
where $N$ is the total number of samples in the simulation and the subscript $i$ denotes the output (i.e. 1 for $v$ and 2 for $\rho$). Table \ref{tab: tank comp2} shows the $SSE_i$ according to (\ref{eq:min sse total}) achieved in the tank volume $v$ and output density $\rho$ as well as the total weighted $SSE_{tot}$ according to (\ref{eq:min sse}) over the entire six hour simulation period for each controller acting alone for the same disturbance scenario and on the same plant with the 10\% gain uncertainty in the input water flow-rate $q_w$. It can be seen that the difference in performance in terms of the density $\rho$ between the different controllers is not very large with a 5\% difference between the best and worst controllers.

\renewcommand{\arraystretch}{1.3}
\begin{table}[h]
	\centering
	\caption{Controller performance indices with the platform and for each individual controller acting alone}
	\label{tab: tank comp2}
	\begin{tabular}{l c c c c c c}
		\hline
		&  $SSE_{\rho}$ & $SSE_{v}$  & $SSE_{tot}$  \\
		No Platform ($\mathbf{G}_{c\_local}$)	& 0.6909 & 62.12  & 0.7530 \\ 
		With Platform  							& 0.6757 & 10.42 & 0.6861   \\ 
		Inverse Based Controller ($\mathbf{G}_{c1}$)  & 0.6714 & 25.00 & 0.6964 \\
		Linear MPC Controller ($\mathbf{G}_{c2}$) 	& 0.6717 & 7.337 & 0.6790 \\
		Nonlinear MPC Controller ($\mathbf{G}_{c3}$) 	& 0.6711 & 7.605 & 0.6787 \\
	\end{tabular}
\end{table}
There is a much bigger discrepancy in terms of the volume $v$, with the local controller $\mathbf{G}_{c\_local}$ performing extremely poorly (as loop interactions are not taken into account) and the MPC controllers performing the best by a large margin. In terms of the weighted overall performance, the nonlinear MPC controller $\mathbf{G}_{c_3}$ is the best, outperforming the linear MPC controller $\mathbf{G}_{c_2}$ by only a narrow margin. It can also be seen that a drastic improvement in performance has been achieved through the use of the platform providing the plant access to the better performing controllers.

It is evident from Fig. \ref{fig:c_2} that the selector has chosen the linear MPC controller ($\mathbf{G}_{c2}$) for the majority of the time, where Table \ref{tab: tank comp2} indicates that the nonlinear MPC controller ($\mathbf{G}_{c3}$) would have performed better (although very marginally). This can be explained with the help of Table \ref{tab: tank comp3} which shows the $SSE_{tot}$ determined by the platform from the simulation of each controller on the accurate plant model. 

It can be seen in this table that the linear MPC controller ($\mathbf{G}_{c2}$) marginally outperforms the nonlinear MPC controller ($\mathbf{G}_{c3}$), with the performance measure almost equal over some evaluation periods. This can be attributed to the fact that the disturbance estimates in the observer play a major role in the disturbance rejection properties of the controller, and since both controllers use the same linear observer, their performance is similar. 

There is therefore a minor discrepancy in these results between the theoretical performance determined by the selector and the results of the actual simulation in the presence of plant model mismatch. This presents a potential difficulty for this platform philosophy in that plant model mismatch is always present in practice and would be difficult to factor into the decision made by the selector, but in this case, the effect is negligible as the performance of the linear and nonlinear MPC controllers differ by a negligible amount.

\renewcommand{\arraystretch}{1.3}
\begin{table*}[h!]
	\centering
	\caption{Controller performance indices for each individual controller over each evaluation horizon $N_e$. \textcolor{green!55!blue}{The best performance in each row is colour-coded.} }
	\label{tab: tank comp3}
	\begin{tabular}{c c c c c c c c}
		\hline
		$N_e$ (hours)  & $\mathbf{G}_{c\_local}$ (PI) & $\mathbf{G}_{c1}$ (Mod. Inv.)&
		$\mathbf{G}_{c2}$ (LMPC) &	$\mathbf{G}_{c3}$ (NMPC)	 \\
		0 - 0.5 & 1.728e-3	 &\textcolor{green!55!blue}{6.693e-4}   & 7.822e-4 & 9.060e-4	\\
		0.5 - 1 & 1.938e-2	& 6.190e-3	& \textcolor{green!55!blue}{5.370e-3}	&5.443e-3	\\
		1 - 1.5 & 1.844e-2  & 6.154e-3 & \textcolor{green!55!blue}{5.572e-3} &	5.612e-3	\\
		1.5 - 2 & 7.031e-3  & 3.247e-3 &\textcolor{green!55!blue}{3.207e-3}	 &	3.759e-3	\\
		2 - 2.5 & 1.469e-2  & 4.016e-3 & \textcolor{green!55!blue}{3.567e-3} & 3.573e-3 		\\
		2.5 - 3 & 7.391e-1  &	4.064e-1 & \textcolor{green!55!blue}{4.038e-1}  & 4.057e-1	\\
		3 - 3.5 & 4.068e-1  &\textcolor{green!55!blue}{2.275e-1}	 & 2.369e-1 & 2.369e-1		\\
		3.5 - 4 & 4.642e-3  &	2.670e-3 & \textcolor{green!55!blue}{2.595e-3} & 2.960e-3	\\
		4 - 4.5 & 1.132e-2  & 2.872e-3  & \textcolor{green!55!blue}{2.645e-3} & 2.645e-3		\\
		4.5 - 5 & 1.647e-2  & 5.079e-3 &\textcolor{green!55!blue}{4.479e-3}	 & 4.479e-3		\\
		5 - 5.5 & 1.416e-2  & 4.658e-3 &\textcolor{green!55!blue}{4.206e-3}	 & 4.206e-3		\\
		5.5 - 6 & 1.564e-3  & 4.941e-4 &\textcolor{green!55!blue}{6.255e-4}	 & 6.278e-4		\\
	\end{tabular}
\end{table*}

\section{Conclusion}

This paper proposes an innovative approach for the advanced control of an industrial process via an automation cloud platform. A case was made that it possible and beneficial for users of automation technology to have multiple vendors compete to control an industrial process via an automation cloud platform which facilitates the interaction between APC vendors and the process. A selector, which forms part of the platform, was used to determine the best controller for a process for any given time period. 
								
An overview of platform businesses was given with a particular focus on platforms aimed at industry. A brief analysis was given of issues that need to be addressed to make APC via an industrial automation cloud platform practically viable. These included what process information to provide to APC vendors, continuous evaluation of controllers even when not in control of the process, bumpless transfer, closed-loop stability, constraint handling, and platform security and trust.

A case study was then given of competing APCs via an industrial automation cloud platform. The process used in the study is a surge tank from a bulk tailings treatment plant for which a detailed plant model was derived and analyzed. It was shown that adequate control of the plant cannot be achieved by rudimentary decentralised PI control owing to plant interactions. APC via a platform that facilitates the competition of three vendors for control of the process, was shown to provide the plant access to a superior control without the need for directly procuring the services of an exclusive vendor.

%%%%%%%%%%%%%%%%%%%%%%%%%%%%%%%%%%%%%%%%%%%%%%%%%%%%%%%%%%%%%%%%%%%%%%%%%%%%%%%%%%%%%%%%%%%%%%%%%%%%%%%%%%%%%%%

\appendix

\section{Surge Tank Model Input/Output Controllability Analysis} \label{Appendix: IO cntrl}

A brief input-output controllability analysis is conducted on the linearised model based on that presented in \cite{Skogestad2005}. This is done in order to ascertain if it is possible to keep the outputs $\mathbf{y}$ within specified bounds from the reference $\mathbf{r}$ using the available actuator authority $\mathbf{u}$, in the presence of disturbances $\mathbf{d}$. This analysis includes the following:
\begin{enumerate}
	\item Modal/state controllability and observability analysis.
	\item Determine the multi-variable poles and zeros of the plant.
	\item Determine the relative gain array of the plant.
	\item Scale the plant so that all (scaled) inputs, outputs and disturbance are confined to values between zero and one and plot the singular values of the scaled plant model $\tilde{\mathbf{G}}_p$ as a function of frequency. It is desired that the minimum singular value of the scaled plant model $\tilde{\mathbf{G}}_p$ be large at frequencies where control is needed. Large singular values in the disturbance model$\tilde{\mathbf{G}}_d$ indicate the need for control action to reject the disturbances.
	\item  Determine the elements of the matrix product $\tilde{\mathbf{G}}_p^{-1}\tilde{\mathbf{G}}_d$. If all elements are less than one over all frequencies then perfect disturbance rejection is possible.
\end{enumerate}

\subsection{Modal/State Controllability and Observability}\

The controllability and observability matrices:
\begin{equation}
\label{eq:cntrblty}
\mathit{C} = \begin{bmatrix}
\mathbf{B} && \mathbf{AB} && \mathbf{A}^2\mathbf{B} && \cdots &&
\mathbf{A}^{n-1}\mathbf{B} 
\end{bmatrix}
\end{equation}
\begin{equation}
\label{eq:obsrvblty}
\mathit{O} = \begin{bmatrix}
\mathbf{C} \\ \mathbf{CA} \\ \mathbf{C}\mathbf{A}^2 \\ \vdots \\
\mathbf{C} \mathbf{A}^{n-1} 
\end{bmatrix}
\end{equation}

applied to (\ref{eq:lin_model_ss}) yields:
\begin{equation}
\mathit{C} = \begin{bmatrix}
1 && 1 && 0 && 0 \\
0.01 && -0.04 && -0.75 && 3 
\end{bmatrix}
\end{equation}
and 
\begin{equation}
\mathit{O} = \begin{bmatrix}
1 && 0 \\ 0 && 1 \\ 0 && 0 \\ 0 && -75 
\end{bmatrix}
\end{equation}

Both matrices have a rank of two, thus the system is completely controllable and observable \citep{Sontag1998}.

The $\mathbf{A}$ matrix in (\ref{eq:lin_model_ss}) has two eigenvalues and corresponding right eigenvectors given by:
\begin{equation}
\lambda_1  = 0 , \quad	\lambda_2  = -75
\end{equation}
\begin{equation}
\mathbf{v}_1  = \begin{bmatrix} 1\\0 \end{bmatrix}	, \quad
\mathbf{v}_2  = \begin{bmatrix} 0\\1 \end{bmatrix}.	
\end{equation}

These two modes correspond directly with the two states of the system (i.e. $v$ and $\rho$). It can therefore be seen from inspection of the $\mathbf{B}$ matrix in (\ref{eq:lin_model_ss}) that both states are each influenced by both inputs. The two inputs $u_1$ and $u_2$ (corresponding to $q_i$ and $q_w$ respectively) have an equal influence on the rate of change of the first state $x_1$ (corresponding to $v$) of unity gain, while they effect the rate of change of the second state $x_2$ by factors of 0.01 and -0.04 respectively. Intuitively this makes sense, as the tank volume $v$ is equally influenced by changes in the flow rates $q_w$ and $q_i$.  Similarly, increasing the flow rate of product $q_i$ would cause an increase in the output density $\rho$ (provided the input density $\rho_i$ is greater than the nominal $\rho^*$), while increasing the flow rate of water $q_w$ would cause a decrease in the output density.   

In terms of observability, the two outputs of the system correspond directly with each state (the $\mathbf{C}$ matrix in (\ref{eq:lin_model_ss}) is the identity matrix), which in turn are equivalent to the two modes of the system. From an observability perspective, it can be said that each mode is independently observable in each corresponding output.

\subsection{Multi-variable Poles and Zeros}\

The plant poles are defined as the eigenvalues of the $\mathbf{A}$ matrix of the plant \citep{Skogestad2005}. The plant has two poles, one at the origin corresponding to the eigenvalue $\lambda_1$ of 0 and another at -75 corresponding to the second eigenvalue $\lambda_2$. This is also evident from the transfer function matrix representation of the plant given in (\ref{eq:lin_model_tf}). It is thus evident that the plant has a pole at the origin which corresponds with the tank volume, which by nature is an integrating process. 

A plant model has a multi-variable zero if, at some frequency or for some frequency range, the transfer function matrix loses rank \citep{Skogestad2005}. Thus for the transfer function matrix $\mathbf{G}_p$ in (\ref{eq:lin_model_tf}), the normal rank of the matrix is two, and there is no frequency $\omega$ where substitution of $s = j\omega$ will cause the matrix to lose rank. The plant model therefore does not have any multi-variable zeros. 

\subsection{Relative Gain Array}\

The relative gain array (RGA) of a non-singular square plant model is defined as:
\begin{equation}
\label{eq:RGA}
RGA(\mathbf{G}_p) = \mathbf{G}_p \times (\mathbf{G}_p^{-1})^T
\end{equation}
where $\times$ indicates the element by element multiplication of the matrix \citep{Skogestad2005}. The relative gain array provides an indication of the input-output controllability of a plant in a number of ways. Firstly, plants that produce large RGA elements are fundamentally difficult to control due to sensitivity to uncertainty. Secondly the RGA indicates the preferred input output pairing if decentralised control is to be used, with the preferred pairing chosen corresponding to the RGA values closest to one, and avoiding pairings which correspond to negative RGA elements \citep{Skogestad2005}. 

Applying (\ref{eq:RGA}) to the plant model in (\ref{eq:lin_model_tf}) gives:
\renewcommand{\arraystretch}{1}
\begin{equation}
\label{eq:RGA_plant}
\begin{split}
\begin{bmatrix}
\dfrac{1}{s}&&\dfrac{1}{s}\\[12pt]
\dfrac{0.01}{s+75} && \dfrac{-0.04}{s+75}
\end{bmatrix} \times &
\begin{bmatrix}
0.8s&&0.2s\\
20(s+75) && -20(s+75)
\end{bmatrix}
\\	
= & \begin{bmatrix}
0.8&&0.2\\
0.2 && 0.8
\end{bmatrix}
\end{split}
\end{equation}

It can be seen from (\ref{eq:RGA_plant}) that the diagonal elements are closer to one than the off diagonal elements, thus for decentralised control, input output pairing should be done along the diagonal elements. For the plant in question, the flow rate of the product into the tank would be best used to control the tank volume, where the flow rate of water into the tank would best be used to control the density of the slurry in the tank. The magnitude of the RGA elements also provide an indication of the interaction between inputs and outputs, where a value of zero would indicate no interaction and a value close to one indicating that only the corresponding input effects the corresponding output. A completely decoupled, diagonal plant will therefore have an RGA matrix equal to the identity matrix \citep{Skogestad2005}. In this case, the RGA is not equal to the identity matrix but does indicate a relatively good level of decoupling between inputs and outputs. It is also evident that uncertainty will not be much of an issue in controlling the plant due to the small value of the RGA elements.

\subsection{Scaled Model Singular Values}\

The scaled plant and disturbance models $\tilde{\mathbf{G}}_p$ and $\tilde{\mathbf{G}}_d$ are determined as:
\begin{equation}
\label{eq:scaling}
\begin{split}
\mathbf{D}_y\tilde{\mathbf{y}} &= \mathbf{G} \mathbf{D}_u\tilde{\mathbf{u}}	+ 
\mathbf{G}_d \mathbf{D}_d\tilde{\mathbf{d}} \\
\tilde{\mathbf{y}} &=  \mathbf{D}_y ^{-1}\mathbf{G} \mathbf{D}_u\hat{\mathbf{u}}	+ 
\mathbf{D}_y ^{-1}\mathbf{G}_d \mathbf{D}_d\tilde{\mathbf{d}} \\
\tilde{\mathbf{y}} &=  \tilde{\mathbf{G}}_p\hat{\mathbf{u}}	+ 
\tilde{\mathbf{G}}_d \tilde{\mathbf{d}}
\end{split}
\end{equation}
where $\tilde{\mathbf{y}}$, $\tilde{\mathbf{u}}$ and $\tilde{\mathbf{d}}$ are the scaled output, input and disturbance which take on values between 0 and 1. The scaling matrices $\mathbf{D}_y$, $\mathbf{D}_u$ and $\mathbf{D}_d$ are diagonal matrices with the elements taken as the minimum value of the corresponding signal for the outputs and inputs, and the maximum value of the corresponding signal for the disturbances. These values are determined from Table \ref{tab: tank par} in deviation variable form.

The scaled plant and disturbance models $\tilde{\mathbf{G}}_p$ and $\tilde{\mathbf{G}}_d$ are found to be:
\begin{equation}
\label{eq:scaled_plant_model}
\tilde{\mathbf{G}}_p = \begin{bmatrix}
\dfrac{85.71}{s}&&\dfrac{21.43}{s}\\[12pt]
\dfrac{60}{s+75} && \dfrac{-60}{s+75}
\end{bmatrix}
\end{equation}
\begin{equation}
\label{eq:scaled_dist_model}
\tilde{\mathbf{G}}_d = \begin{bmatrix}
0 \\[6pt] \dfrac{300}{s+75}
\end{bmatrix}
\end{equation}

\begin{figure}[h!]
	\centering
	\includegraphics[width=9cm]{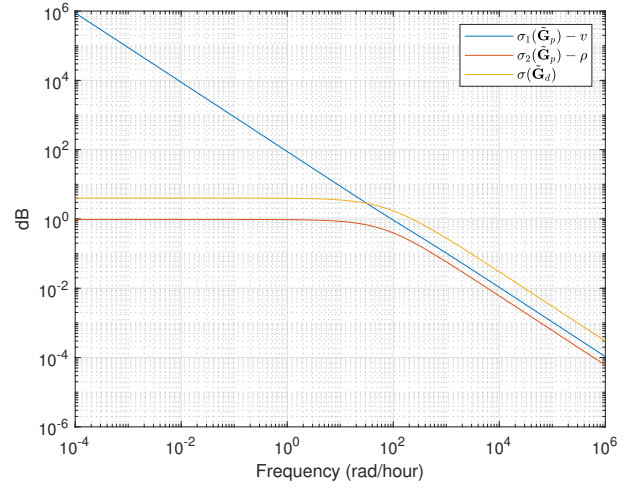}
	\caption{Scaled plant model singular values vs. frequency}
	\label{fig:plant_svd}
\end{figure}

Fig. (\ref{fig:plant_svd}) shows plots of the singular values of the scaled plant and disturbance models in (\ref{eq:scaled_plant_model}) and (\ref{eq:scaled_dist_model}) versus frequency.
It can be seen from Fig. \ref{fig:plant_svd} that at low frequencies the minimum singular value takes on a value of one, while the maximum singular value increases indefinitely with decreasing frequency. This can be explained intuitively as the maximum singular value corresponds with the volume of the tank $v$ (i.e. integrators in the plant model) while the smaller singular value corresponds with the density $\rho$. In this case, any tank level can be achieved due to the integrating nature of the process, hence the larger singular value increases as frequency decreases. The minimum singular value is approximately equal to 1 at frequencies where control is necessary and the density $\rho$ is thus functionally controllable.

At frequencies where the maximum singular value of the disturbance model (in this case there is only one singular value) is less than one, the output will remain within specified limits for all possible values of the disturbance variable, thus large singular values in the disturbance model indicate a need for control in order to reject disturbances at these frequencies. It can be seen in Fig. \ref{fig:plant_svd} that the singular value of the disturbance model does exceed one. At low frequencies, the steady-state value is $300/75=4$ and thus control is needed to reject disturbances at frequencies below approximately 200 $rad/hour$.

\subsection{Disturbance Rejection Capabilities}\

In order to achieve perfect disturbance rejection, the maximum singular value of the matrix product $\mathbf{G}^{-1}\mathbf{G}_d$ needs to be less than one over all frequencies \citep{Skogestad2005}. The analytical calculation of $\mathbf{G}^{-1}\mathbf{G}_d$ gives:
\begin{equation}
\label{eq:GinvGd}
\begin{split}
\mathbf{G}^{-1}\mathbf{G}_d & = 
\begin{bmatrix}
0.00933s&&0.00333(s+75)\\[12pt]
0.00933s && -0.01333(s+75)
\end{bmatrix}
\begin{bmatrix}
0 \\ \dfrac{300}{s+75}
\end{bmatrix}  \\
& = \begin{bmatrix}
1 \\ 4
\end{bmatrix}
\end{split}
\end{equation}
which is constant over all frequencies. It is thus evident that perfect disturbance rejection is not possible. This can be intuitively deduced from the system constraints given in Table \ref{tab: tank par}. The disturbance variable $\rho_i$ can vary between 1 and 2 $t/m^3$, thus if the set-point for the density $\rho$ is fixed at the nominal value of 1.4 $t/m^3$, it will not be possible to maintain this density if the density of the feed falls below 1.4 $t/m^3$, in which case the water feed will need to be shut off ($q_w=0$) and the input feed $q_i$ flow rate set to the nominal output feed flow rate of 750 $m^3/hour$ to keep the level constant. Perfect disturbance rejection is therefore not possible for the range of input densities given in Table~\ref{tab: tank par}.

%It is worth noting that there is a certain amount of inaccuracy between the linear and nonlinear models in terms of the change in the feed flow rate $q_i$ and the water flow rate $q_w$ required to reject a given change in the feed density $\rho_i$. For example if there is a increase of 0.5 $t/m^3$ in $\rho_i$ to its maximum value of 2 $t/m^3$, the required increase and decrease in the feed flow rate $q_i$ and the water flow rate $q_w$ when calculated according to the linear model is double the result when calculated according to the nonlinear model. A similar calculation for an increase in 0.1 $t/m^3$ in $\rho_i$ yields a 20\% difference in the result between the nonlinear and linear models. It is thus evident that nonlinearities in the model have an effect on the input magnitude required to reject a given disturbance with the magnitude of the effect increasing with the magnitude of the disturbance.

\end{document}